\documentclass[usenatbib]{mnras}
\usepackage{newtxtext,newtxmath}
\usepackage{tikz}
\usetikzlibrary{arrows.meta}

\usepackage[T1]{fontenc}
\DeclareRobustCommand{\VAN}[3]{#2}
\let\VANthebibliography\thebibliography
\def\thebibliography{\DeclareRobustCommand{\VAN}[3]{##3}\VANthebibliography}
\usepackage{xcolor}

\usepackage{graphicx}	% Including figure files
\usepackage{tikz,tikz-3dplot}
\usetikzlibrary{3d,quotes,angles}
\usepackage{pst-3dplot}% http://ctan.org/pkg/pst-3dplot
\usepackage[thinc]{esdiff}
\usepackage{braket}

\usepackage{amsmath}	% Advanced maths commands
\usepackage{mathtools}
\usepackage{hyperref}
\hypersetup{colorlinks=true,
            citecolor=blue,
            linkcolor=blue}

\usepackage{chngcntr}
\usepackage{newtxtext,newtxmath}

\usepackage{varwidth}
\usepackage{soul}

\newcommand{\msun}{\mathrm{M}_{\odot}}	% Solar mass
\newcommand{\hillradius}{r_{\rm H}}

\newcommand{\mBH}{M_{\rm BH}}
\newcommand{\ph}{p_{\rm 1H}}
\newcommand{\fform}{f_{\rm form}}

\newcommand{\mbin}{M_{\rm bin}}
\newcommand{\mSMBH}{M_\bullet}

\defcitealias{Rowan2022}{Paper~I}
\defcitealias{Rowan2023}{Paper~II}
\begin{document}

%%%%%%%%%%%%%%%%%%% TITLE PAGE %%%%%%%%%%%%%%%%%%%

% Title of the paper, and the short title which is used in the headers.
% Keep the title short and informative.
\title[Merger rates in AGN]{Black Hole Merger Rates in AGN: contribution from gas-captured binaries}

% The list of authors, and the short list which is used in the headers.
% If you need two or more lines of authors, add an extra line using \newauthor
\author[C. Rowan et al.]{
Connar Rowan$^{1,2}$\thanks{E-mail: connar.rowan@nbi.ku.dk}, Henry Whitehead$^{3}$ and Bence Kocsis$^{1,4}$
\\
$^{1}$Rudolf Peierls Centre for Theoretical Physics, Clarendon Laboratory, University of Oxford, Parks Road, Oxford, OX1 3PU, UK
\\
$^{2}$Niels Bohr International Academy, The Niels Bohr Institute, Blegdamsvej 17, DK-2100, Copenhagen , Denmark 
\\
$^{3}$Department of Physics, Astrophysics, University of Oxford, Denys Wilkinson Building, Keble Road, Oxford OX1 3RH, UK
\\
$^{4}$St Hugh's College, St Margaret's Rd, Oxford, OX2 6LE, UK
\\
}
% These dates will be filled out by the publisher
%\date{Accepted XXX. Received YYY; in original form ZZZ}

% Enter the current year, for the copyright statements etc.
%\pubyear{2015}

%\label{firstpage}
\date{\today}


\maketitle

\begin{abstract}
It has been suggested that merging black hole (BH) binaries in active galactic nucleus (AGN) discs formed through two-body scatterings via the ``gas capture'' process may explain a significant fraction of BH mergers in AGN and a non-negligible contribution to the observed rate from LIGO-VIRGO-KAGRA. We perform Monte Carlo simulations of BH and binary BH formation, evolution and mergers across the observed AGN mass function using a novel physically motivated treatment for the gas-capture process derived from hydrodynamical simulations of BH-BH encounters in AGN and varying assumptions on the AGN disc physics. The results suggest that gas-captured binaries could result in merger rates of $0.73-7.1$Gpc$^{-3}$yr$^{-1}$. Most mergers take place near the outer boundary of the accretion disk, but this may be subject to change when migration is considered. The BH merger rate in the AGN channel in the Universe is dominated by AGN with supermassive BH masses on the order of $\sim10^{7}\msun$, with $90\%$ of mergers occurring in the range $\sim10^{6}\msun-10^{8}\msun$. The slope of the merging mass distribution is flatter than the initial BH mass power law by a factor $\Delta \xi=1.1-1.2$, as larger BHs can align with the disc and successfully form binaries more efficiently. Similarly, the merging mass ratio distribution is flatter, therefore the AGN channel could easily explain the high mass and unequal mass ratio detections such as GW190521 and GW190814. When modelling the BH binary formation process using a simpler dynamical friction treatment, we observe very similar results, where the primary bottleneck is the alignment time with the disk.
We find the most influential parameters on the rates are the anticipated number of BHs and their mass function. We conclude that AGN remain an important channel for consideration, particularly for gravitational wave detections involving one or two high mass BHs.
\end{abstract}

% Select between one and six entries from the list of approasssssssssaasved keywords.
% do not make up new ones.
\begin{keywords}
binaries: general -- transients: black hole mergers -- galaxies: nuclei -- Hydrodynamics -- Gravitational Waves
\end{keywords}

%%%%%%%%%%%%%%%%%%%%%%%%%%%%%%%%%%%%%%%%%%%%%%%%%%

%%%%%%%%%%%%%%%%% BODY OF PAPER %%%%%%%%%%%%%%%%%%
\section{Introduction}
Following the first gravitational wave (GW) detection in 2015 \citet{LIGO2016}, we have been able to observationally detect BHs through both electromagnetic \citep[e.g][]{HMXB_2006,LMXB_2007,Hailey2018,Thompson2018,EHT2019,Akiyama2022,Gaia_2024}, and GW observations \citep[e.g][]{LIGO2019,LIGO2020a,LIGO2020b,LIGO2020c,LIGO2020d}, with over 170 more sources anticipated from the LIGO public alert list for the O4 observing run\footnote{see \url{https://gracedb.ligo.org/superevents/public/O4/} for the current number of alerts.}. The merging of BHs from gravitational wave emission requires the binary black hole (BBH) system to have an extremely small separation for the system to merge within a Hubble time ($\sim$0.1au for two 10$\msun$ BHs). As the separation of stellar binaries is typically far greater than this limit \citep[][]{Opik1924,Tokovinin2000}, what astrophysical mechanism(s) can bring BHs to such low separations is still an open question. Several possible explanations have been suggested: isolated stellar binary evolution and common envelope evolution \citep[e.g][]{Lipunov1997,Belczynski2010,Belczynski2016,Dominik2012,Dominik2013,Dominik2015,Tagawa2018,Mapelli2021}, three body scatterings in nuclear and globular star clusters \citep[e.g][]{Mouri2002,Miller2002,Antonini+2012,Rodriguez+2015,Rodriguez+2016,Samsing2018,DiCarlo2020,Liu2021}, BBH mergers driven by gas in globular clusters \citep{Rozner2022} and in dense AGN discs \citep[e.g][]{Bartos2017,Stone2017,Tagawa2020,Li2021,Li_and_lai_2022,Li_Dempsey_Lai+2022,Rowan2022,Rowan2023,Whitehead2023,Whitehead2023_novae,McKernan2020_AGNmontecarlo,McKernan+2020,Delfavero2024}. 

AGN provide a favourable environment for both BBH formation and mergers, due to the dense gaseous accretion disc orbiting the super-massive black hole (SMBH) at their centre \citep{Tagawa2020}. Dynamical/accretion drag on objects crossing through the AGN disc can embed them within the geometrically thin disc for a considerably high number density of BHs \citep{Bartos2017,Fabj2020}. These BHs can then encounter one another and form stable binaries by dissipating their relative two-body energy via a complex interaction with the surrounding gas. The efficiency of this `gas-assisted' binary formation mechanism has recently been validated by hydrodynamical simulations in our own works \citep[e.g][]{Rowan2022,Rowan2023,Whitehead2023,Whitehead2023_novae} and in e.g. \citet{Li_Dempsey_Lai+2022}, see also the analytical studies of \citet{DeLaurentiis2022} and \citet{Rozner+2022}. \citet{Dodici2024} directly compared across previous simulation studies, finding good agreement for the island of parameter space in gas density and encounter impact parameter that leads to binary formation. 

The evolution of a disc embedded binary is an open problem. In \citet{Rowan2022}, we demonstrated that binaries can inspiral on short timescales ($\lesssim10^{4}$yr), orders of magnitude shorter than the typical lifetime of an AGN, $t_\mathrm{AGN}=10^{7}$yr. The merger can be especially rapid for retrograde binaries (retrograde meaning the binary orbits counter to the orbit about the SMBH), due to enhanced accretion and gravitational torques from the gas that remove angular momentum. This picture is broadly consistent across other simulation studies with varying methodoligies \citep[e.g][]{Baruteau2011,Secunda2019,Li_and_lai_2022,Li_Dempsey_Lai+2022,Li_and_Lai_2022_windtunnel_II,Li_and_Lai_windtunnel_III_2023}. For prograde binaries that are highly circular and at larger separations, binaries have in some cases been found to out-spiral when an isothermal hydrodynamic treatment is used in 2D \citep[e.g][]{Li2021}. 

With the ever increasing number of BBH merger detections from LIGO-VIRGO-KAGRA (LVK), we continually constrain the merger rate of these systems. Population studies and semi-analytic works \citep[e.g][]{Bartko2009,Antonini+2016,Belovary2016,Tagawa2020,McKernan2020,ford2022_AGNrates,Vaccaro2024} that predict the rate of mergers in AGN tend to suffer from a weakly motivated or highly simplified treatment of binary formation in gas, which naturally affects the anticipated rates. Arguably the most detailed of these studies is the 1D N-body simulations of \citet{Tagawa2020,Tagawa2020_spin,Tagawa+2021_eccentricity}. In that study a plethora of physical effects were accounted for, including: radial migration, binary-single interactions, merger kicks and repeated mergers among many others. They found that gas-captured binaries make up the majority ($>85$\%) of merging binaries in the AGN. If the binaries formed through the gas-formation mechanism do indeed represent the primary contribution to the BH mergers in AGN, it is vital that we accurately model the process. \citet{Tagawa2020} assumed a binary was formed if the deceleration timescale from dynamical friction \citep[e.g][]{Ostriker1999} was smaller than the binaries' Hill sphere crossing time. The validity of applying the Ostriker formula to this scenario is dubious, as the formalism assumes a uniform density and motion of the gas, whereas the embedded BHs will have their own very dense and rotating circum-single discs. This has been affirmed by detailed hydrodynamical studies \citep[e.g][]{Li_Dempsey_Lai+2022,Rowan2022,Whitehead2023,Rowan2023}.

In this work, we build on two of our previous works \citet{Rowan2023} and \citet{Whitehead2023} where a detailed capture criterion was derived from high resolution simulations of BH-BH scatterings. We apply this physically motivated and numerically verified criterion to a semi-analytic model of BHs in an AGN disc to test how the merger rates in the AGN channel change with the improved treatment of the gas-assisted binary formation mechanism, with a direct comparison to the simpler dynamical-friction-based formation criterion. We discuss the semi-analytic model and the inclusion of the formation criterion in section \ref{sec:methods} and present our results in section \ref{sec:results} before concluding in section \ref{sec:conclusions}. In section \ref{sec:gas_dissipation}, we describe the process of integrating the binary formation model from the hydrodynamical simulations to the Monte Carlo simulations of this work. References to equations in these papers are denoted explicitly in square brackets (e.g Eq. [27]) for clarity.
\section{Methods}
\label{sec:methods}
The core component of the AGN channel is the SMBH's accretion disc. Here we outline the properties of the disc, how the BHs become embedded into the disc, how gas can lead to formation and finally how formed binaries can be driven to merge.
\subsection{Disc setup}
\label{sec:disc_setup}
We consider the AGN disc model of \citet{Sirko2003} (hereafter SG discs). To observe the dependence on the mass of the AGN, we assume values for the SMBH mass $\mSMBH$ in the range $\mSMBH=10^{5}\msun-10^{9}\msun$ (consistent with the observed AGN mass range, e.g \citealt{Greene_Ho2007,Greene2009}) in uniform log-space, using 35 values. 

The kinematic viscosity of the gas $\nu$, which determines the angular momentum flow in SG discs, is related to the sound speed $c_\mathrm{s}$ and disc scale height $H$ via
\begin{equation}
    \nu=\alpha\beta^{b} c_\mathrm{s} H\,. 
    \label{eq:viscosity}
\end{equation}
Here, $\alpha$ is the viscosity constant \citep[][]{Shakura1973} and $\beta\equiv P_\mathrm{gas}/(P_\mathrm{gas}+P_\mathrm{rad})$ is the ratio of the gas pressure $P_\mathrm{gas}$ to the total pressure, which includes the radiation pressure $P_\mathrm{rad}$. The parameter $b=\{0,1\}$ acts as the switch between an $\alpha$-disc (b=0) and a $\beta$-disc ($b=1$), see \citet{Haiman2009} for more on this viscosity treatment. We consider both the $\alpha$-disc and $\beta$-disc treatment of the viscosity for our models of the AGN disc. We tack on the $\alpha$, $\beta$ labels to the SG abbreviation, such that 'SG$\alpha$' corresponds to a Sirko-Goodman disc with the $\alpha$ viscosity treatment and 'SG$\beta$' corresponds to a SG disc with the $\beta$ viscosity treatment. In the SG model, the free parameters are the mass of the SMBH $\mSMBH$, the mean molecular mass $\mu_\mathrm{mol}$, the Eddington ratio $\epsilon=L/L_{\rm Edd}$ (where $L_\mathrm{Edd}$ is the Eddington luminosity) and viscosity constant $\alpha$. Inline with our previous work \citep{Rowan2022,Rowan2023}, we set $\alpha=0.1$,  $\mu_\mathrm{mol}=0.6$ and $\epsilon=0.2$. 

We solve the disc equations explicitly, using the \texttt{pAGN} code of \citet{Gangardt2024} to obtain the radial disc profiles of density $\rho$ and sound speed $c_{\rm s}$. As the binary formation criterion depends strongly on the gas density (see Sec. \ref{sec:formation_probability} below), it is essential to model the AGN discs in an accurate manner\footnote{We fix a minor error in \texttt{pAGN} where the molecular mass is fixed at unity in the ideal gas law.} that does not depend on being in a single disc regime (i.e the gas pressure dominated regime assumed in \citealt{Rowan2022}). We model the discs out to where the Toomre $Q$ parameter predicts gravitational instability, which occurs when
\begin{equation}
    Q\equiv\frac{c_{\rm s} \Omega}{\pi G\Sigma}\lesssim 1\,,
    \label{eq:Toomre}
\end{equation}
where $\Omega=\sqrt{G\mSMBH/R^{3}}$ is the angular frequency of a Keplerian orbit, $G$ is the gravitational constant and $\Sigma$ is the gas surface density. The $R$ value that satisfies Eq. \eqref{eq:Toomre} provides the outer radius of the disc $R_{\rm disc}$ and the upper domain limit for our simulations. Over the range in $\mSMBH$, this value varies minimally around $R\sim0.01$pc. We set the inner $R$ boundary with $R_\mathrm{min}=\max(3R_\bullet,10^{-4}\text{pc})$, where $R_\bullet$ is the Schwarzschild radius.
\subsection{Objects crossing the disc}
We assume all gas-assisted BBH formation and mergers can only occur within the disc, therefore it is necessary to know how many BHs are on orbits that cross the disc. We calculate the number of objects that cross the disc within the simulation domain $R_\mathrm{min}\leq a\leq R_\mathrm{disc}$. The number density of stars $n_{*}$, and by extension their birthed BHs, in terms of the orbital semi-major axis $a$ is taken to be a mass segregated function consistent with the central O-star distribution \citep{Bartko2009,OLeary+2009,Keshet2009}, representing strongly mass segregated equilibrium for the heaviest central objects
\begin{equation}
    \centering
    n_{*}(a)\propto a^{-2.5}\,.
    \label{eq:Radial_dist}
\end{equation}
The distribution is assumed to be spherical with orbital inclinations $i$ sampled uniformly in $\cos{i}$. The eccentricity of the orbits is taken from the thermal distribution $f(e)=2e$. The maximal distance from the SMBH where the local dynamics are dominated by its presence is $R_\mathrm{inf}=GM_\bullet/\sigma^{2}_\bullet$, where $\sigma_\bullet$ is the velocity dispersion of objects in the central nuclear region. Using the $M-\sigma$ relation $\mSMBH=M_{0}(\sigma_\bullet/\sigma_\mathrm{0})^{k}$ \citep{Ferrarese2000,Gebhardt2000,Gultekin2009}, adopting $k=4.384$, $M_{0}=3.097\times10^{8}\msun$ and $\sigma_{\mathrm{0}}=200$km s$^{-1}$ from \citet{Kormendy2013}, this gives
\begin{equation}
    \centering
    R_\mathrm{inf} = \frac{GM_{0}}{\sigma_{0}^{2}}\bigg(\frac{M_{\bullet}}{M_{0}}\bigg)^{0.544}\,.
    \label{eq:R_inf}
\end{equation}
The fraction of BHs crossing the disc is then given by
\begin{equation}
    \centering
    f_\mathrm{cross}=\frac{1}{N_\mathrm{BH}}\underset{a<R_\mathrm{inf}}{\underset{a(1-e)<R_\mathrm{disc}}{\iint}} n(a)f(e)\,4\pi a^{2}da\,de\,,
    \label{eq:Ncross}
\end{equation}
where we integrate over the number of BHs $N_\mathrm{BH}$ within $R_\mathrm{inf}$ that have a periapsis within the outer radial limit of the disc $R_\mathrm{disc}$. The number of BHs sampled in our Monte-Carlo scheme (see \S\ref{sec:monte_carlo}) is $f_\mathrm{cross}N_\mathrm{BH}$.
\subsection{Population sample}
\label{sec:pop_sample}
Based on a standard power law initial mass function (IMF), various models for the BH initial mass function (BIMF) are used in the Monte-Carlo simulations of Sec. \ref{sec:monte_carlo} to monitor their affect on the rates. The stellar population is represented by the initial stellar mass function of \citet{Kroupa2011}
\begin{equation}
    \centering
    \frac{dN_{*}}{dm_{*}}\propto m_{*}^{-\gamma}\,.
    \label{eq:kroupa}
\end{equation}
Here, $m_{*}$ is the zero age main sequence mass of stars with an assumed stellar mass range of $0.1\msun\leq m_{*}\leq 140\msun$ and $N_{*}$ is their abundance. For high stellar masses relevant for forming BHs, the exponent $\gamma$ has a range of $1.7\leq \gamma \leq 2.35$, with the fiducial value taken to be $\gamma = 2.35$, consistent with \citet{Salpeter1955}. The stellar masses are sampled such that the total stellar mass within $R_\mathrm{inf}$ is $2\mSMBH$, in line with \citet{Binney2008,Kocsis2012,Bartos2017}. Three BH initial mass functions are considered, labelled according to their source.

\subsubsection{BIMF 1}
To make an accurate comparison to a highly relevant work, the BIMF of \citet{Tagawa2020} is implemented, corresponding to a solar metallicity \citep[see][]{Belczynski2010,Jermyn2022,Dittmann2023}
\begin{equation}
\frac{m_\mathrm{BH}(m_{*}/\msun)}{\msun} = \left\{
\begin{array}{ll}
      \text{no BH} & m_{*}<20\msun\,, \\
      m_{*}/4 & 20\msun \leq m_{*} < 40\msun \,,\\
      10 & 40\msun \leq m_{*} < 55\msun\,, \\
      m_{*}/13+5.77 & 55\msun \leq m_{*} < 120\msun\,,\\
      15 & 120\msun \leq m_{*} \leq 140\msun\,,
\end{array} 
\right..
    \centering
    \label{eq:BIMF_tagawa}
\end{equation}
with an assumed stellar mass range of $0.1\msun\leq m_{*}\leq 140\msun$. By sampling the stellar masses from Eq. \eqref{eq:kroupa} and applying Eq. \eqref{eq:BIMF_tagawa}, the initial BH mass distribution is obtained. Unlike in \citet{Tagawa2020}, the BH mass function is not evolved here, i.e. only 1st generation mergers are considered.
\subsubsection{BIMF 2}
To also compare with the pre-existing binary rates of \citet{Bartos2017}, we adopt their simplified power law distribution for their binary mass as our BH mass function, 
\begin{equation}
    \frac{dN_\mathrm{BH}}{d\mBH}\propto \mBH^{-\beta}\,,
    \label{eq:BIMF_bartos}
    \centering
\end{equation}
with the equivalent mass range of $5\msun\leq\mbin\leq50\msun$ and exponent $2.0\leq\beta\leq2.5$. The slope of mass distribution reflects the upper bound identified by in \citep{Abbott2019}, which is more consistent with X-ray binary observations \citep[e.g.][]{Ozel2010,Kochanek2015}, which have even steeper dependence. The simplicity of the function allows us to observe how the initial BH mass function compares to the merging mass function, discussed in Sec. \ref{sec:merger_props}. The same number of BHs as derived from the BIMF of \citet{Tagawa2020} is maintained to compare directly the merger rates' dependence on the mass distribution of BHs.
\subsubsection{BIMF 3}
Our only current inference of the BIMF from observations is from X-ray binaries and GW observations. As our third BIMF, we adopt the mass distribution of \citet{Baxter2021}. The function was derived by constraining the mass gap from the BH mass distribution from GW events \citep[see][]{Abbott2021} in tandem with stellar evolution theory, incorporating the predicted mass gap from the GW mass distribution. The function is given by
\begin{equation}
    \centering
    \frac{dN_\mathrm{BH}}{d\mBH}\propto \mBH^{b}\bigg[1+\frac{2a^{2}\mBH^{1/2}(M_\mathrm{BHMG}-\mBH)^{a-1}}{M_\mathrm{BHMG}^{1-\frac{a}{2}}}\bigg]\,,
    \label{eq:BIMF_baxter}
\end{equation}
Where $M_\mathrm{BHMG}$ is the low mass edge of the mass gap. The value of $M_\mathrm{BHMG}$ is set to 47.7$\msun$ and the dimensionless constants are taken to be $a=0.39$ and $b=-2.2$, as suggested in \citet{Baxter2021}. We maintain the same minimum BH mass of $5\msun$ for consistency and use the same number of BHs as the other BIMFs.
%
%\newline\newline

\begin{figure}
    \centering
    \includegraphics[width=8cm]{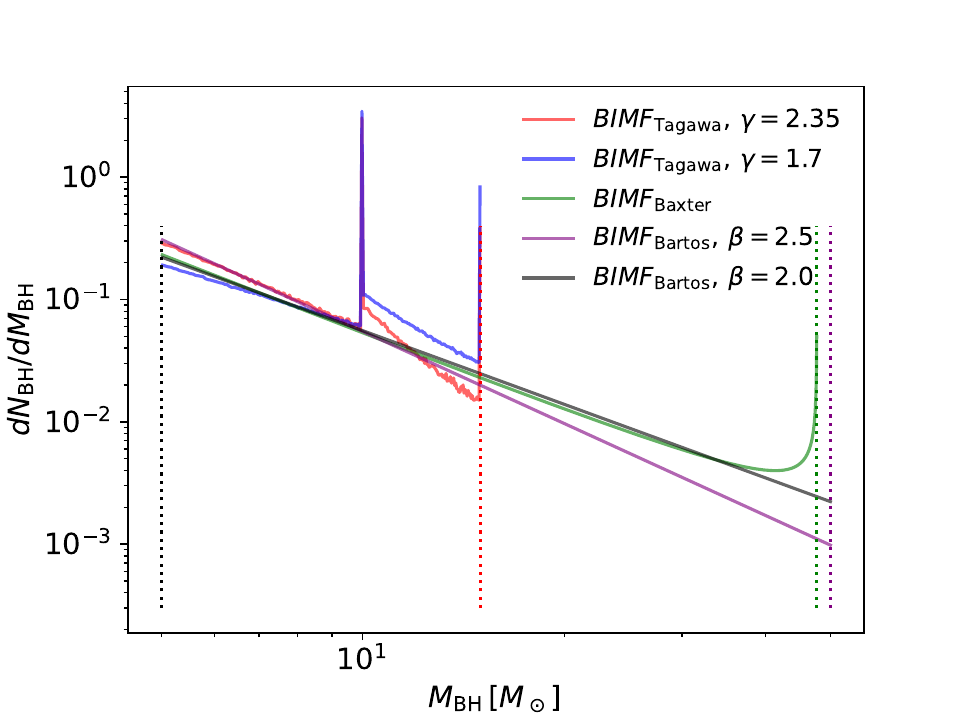}
    \caption{Normalised black hole initial mass functions BIMF$_\mathrm{Tagawa}$, BIMF$_\mathrm{Bartos}$ and BIMF$_\mathrm{Baxter}$ (eqs. \ref{eq:BIMF_tagawa}, \ref{eq:BIMF_bartos} and \ref{eq:BIMF_baxter} respectively). BIMF$_\mathrm{Tagawa}$ is shown for $\gamma=\{1.7,1.35\}$ and  BIMF$_\mathrm{Bartos}$ for $\beta=\{2,2.5\}$. The vertical lines of BIMF$_\mathrm{Tagawa}$ are a result of the $40\msun \leq m_{*} < 55\msun$ and $120\msun \leq m_{*} \leq 140\msun$ conditions of Eq. \eqref{eq:BIMF_tagawa}. The vertical cutoff of BIMF$_\mathrm{Baxter}$, represents the lower boundary of the BH mass gap.}
    \label{fig:IMFs}
\end{figure}

We show the normalised distributions of all IMFs, which we hereafter refer to as BIMF$_\text{Tagawa}$, BIMF$_\text{Bartos}$ and BIMF$_\text{Baxter}$ in Figure \ref{fig:IMFs}. 
Qualitatively, BIMF$_\text{Bartos}$ and BIMF$_\text{Baxter}$ allow for larger BH masses up to $\sim50\msun$, whereas BIMF$_\text{Tagawa}$ produces more  BHs in the range of $10-12\msun$ ($10-15\msun$) for $\gamma=2.35$ (1.7). The spike at $\mBH=10\msun$ comes from the $40-55\msun$ condition of Eq. \eqref{eq:BIMF_tagawa}. As very few stars are formed with masses $120-140\msun$, the anticipated spike at $\mBH=15\msun$ is far less significant.
\subsection{Gas dissipation during the encounter}
\label{sec:gas_dissipation}
\begin{figure}
    \centering
    \includegraphics[width=8cm]{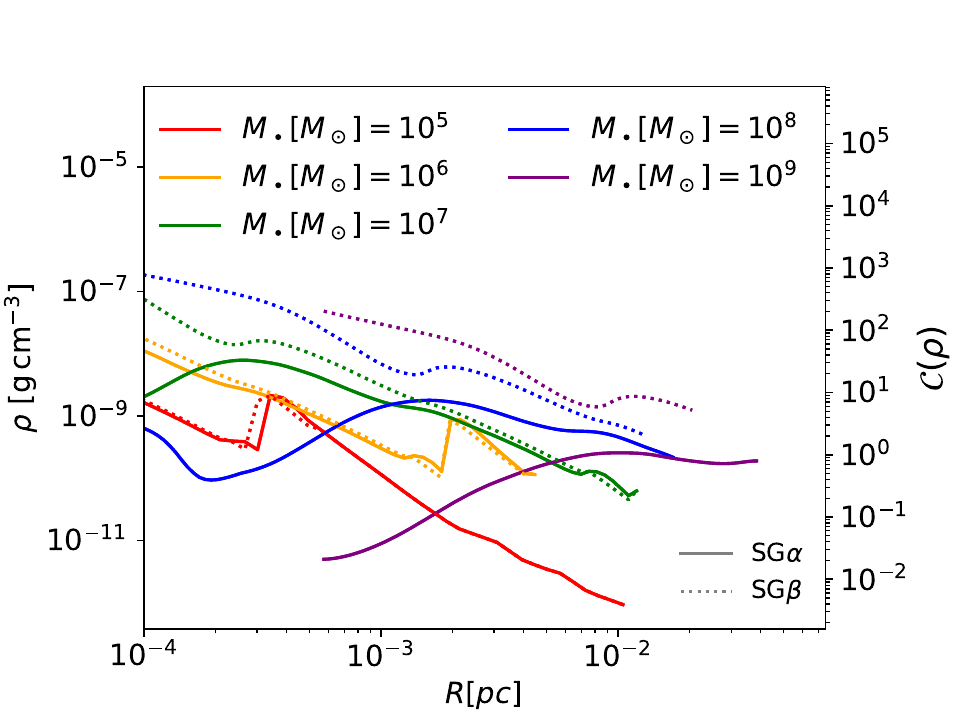}
    \caption{The radial density profiles of the AGN disc generated from \texttt{pAGN} as a function of radial distance in the AGN disc $R$ along side the resulting modification $\mathcal{C}$ to the energy dissipation during encounter (Eqs. \ref{eq:powerlaw_R}-\ref{eq:powerlaw_mod}). Results shown for the SG$\alpha$ and SG$\beta$ disc models with parameters $\mbin=20\msun$ and  $\mSMBH=\{10^{5},10^{6},10^{7},10^{8},10^{9}\}\msun$.}
    \label{fig:C_prof}
\end{figure}

We utilise the semi-analytic prescriptions derived in our previous works \citet{Rowan2023} and \citet{Whitehead2023} to model the energy dissipation of a BH-BH encounter. The orbital energy dissipation $\Delta E_\mathrm{bin}$ from gas drag during the first encounter is assumed to be described by equation [16] in \citet{Rowan2023}. Its form is a power law with the depth of the first periapsis passage $r_\mathrm{per,1}$
\begin{equation}
    \centering
    \Delta E_{\rm{bin}}(r_\mathrm{per,1}) = -x\bigg(\frac{r_\mathrm{per,1}}{\hillradius}\bigg)^{y}|E_\mathrm{H,bin}|\,,
    \label{eq:powerlaw}
\end{equation}
where
\begin{equation}
    |E_\mathrm{H,bin}| = \frac{GM_{\rm{bin}}\mu}{2\hillradius}
\end{equation}
is the absolute orbital energy of the binary in the center-of-mass frame at a separation of one Hill radius, which we define as 
\begin{equation}
    \centering
    \hillradius=R\bigg(\frac{\mbin}{6\mSMBH}\bigg)^{1/3}
    \label{eq:hill}
\end{equation}
 to remain consistent with the parametrisation of \cite{Rowan2023}. We extend the relation to arbitrary densities based on the finding of our other paper \citet{Whitehead2023} that dissipation (normalised to $E_\mathrm{H,bin}$)
%\footnote{under this normalisation, the reduction in the size/mass of the circumsingle discs is accounted for in the $1/\hillradius^{2}$ dependence of $\Delta E_\mathrm{bin}$ as the Hill sphere varies across the AGN disc, assuming they are relatively thin. \BK{This statement is unclear, what does reduction refer to. Why do we need to specify the normalisation for stating the $\rho$ dependence when the normalisation is independent of $\rho$.}} 
scales linearly with the gas density in the Hill sphere $\rho_H$. The two parameters that change $\rho_H$ are the local sound speed and surface density as $\rho_H=\Sigma/(2H)=\Sigma\Omega/c_\mathrm{s}$. We assume $\rho_H$ scales with the ambient density $\rho$ according to the findings of \citet{Whitehead2023}. We modify Eq. \eqref{eq:powerlaw} to account for the changing density in the Hill sphere at different radii in the AGN disc via an additional scaling
\begin{equation}
    \centering
    \frac{\Delta E_{\rm{bin}}(R,r_\mathrm{per,1})}{|E_\mathrm{H,bin}|} = -x\bigg(\frac{r_\mathrm{per,1}}{\hillradius}\bigg)^{y}\mathcal{C}(R,\mSMBH)\,.
    \label{eq:powerlaw_R}
\end{equation}
The function $\mathcal{C}$ is given by
\begin{equation}
 \mathcal{C}(R,\mSMBH) = \bigg(\frac{\rho(R,\mSMBH)}{\rho_0}\bigg)^{w}\,.
 \centering
    \label{eq:powerlaw_mod}
\end{equation}
where $\rho(R,\mSMBH)$ is the ambient density of the disk at the radial location $R$ of the binary.
The normalisation $\rho_0\simeq 6.5\times10^{-10}$g\,cm$^{-3}$ is set by the original simulations in \citet{Rowan2023}. Inline with \citet{Whitehead2023}, we set $w=1$.
The same following methodology is applied from Sec. 3.5 of \citet{Rowan2023} to reconstruct equation [33] using Eq. \eqref{eq:powerlaw_R}. We adopt the fiducial values in \citet{Rowan2023} of $x=1.3\times10^{-4}$, $y=-0.4$\footnote{Though we find in \citet{Rowan2023} that $y=-0.6$ for 3 times the ambient density ($\rho_0\simeq 6.5\times10^{-10}$g\,cm$^{-3}$), it is still within the confidence interval of the error bounds and we retain the same $y=-0.4$ scaling for simplicity. \citet{Whitehead2023} suggests that $y$ is fixed for our fiducial density and below.}. The full capture criterion is then
\begin{equation}
  \bigg[\frac{E_{2\mathrm{H}}}{|E_\mathrm{H,bin}|} < \left(\frac{E_{2\mathrm{H}}}{|E_\mathrm{H,bin}|}\right)_{\rm crit}\bigg]  \wedge \bigg[\ph <0.68\hillradius\bigg]\,,
    \label{eq:cap_criterion}
\end{equation}
\begin{align}
 \left(\frac{E_{2\mathrm{H}}}{|E_\mathrm{H,bin}|}\right)_{\rm crit} &= \underbrace{\mathcal{C}(R,\mSMBH)10^{-1.74\frac{\ph}{\hillradius}-2.52}}_{\text{$|\Delta E_\mathrm{bin}|$ for encounter}} -
 \underbrace{10^{1.31\frac{\ph}{\hillradius}-4.34}}_{\text{max stable $E_\mathrm{bin}$}}
 \,.
 \label{eq:cap_criterion_full}
\end{align}
where $E_{2\mathrm{H}}$ is the binary energy in the center-of-mass frame and $p_\mathrm{1H}=|\hat{\mathbf{v}}_{\rm rel}\times \boldsymbol{\Delta}\mathbf{R}|$ is the impact parameter of the encounter at a separation of $\Delta R = \hillradius$ where the relative velocity unit vector is $\hat{\mathbf{v}}_{\rm rel}$, see equation [30] in \citet{Rowan2023}. This is calculated using a numerical simulation of the single-single scattering process starting from a given set of initial conditions at a separation of $2\hillradius$ without gas. The second term represents the maximum energy for the binaries to remain bound in the simulations of \citet{Rowan2023} for more than 2 encounters (all binaries subsequently hardened after 2 encounters).
We show the value of $\mathcal{C}$ as a function of $R$ for five different $\mSMBH$ according to the \texttt{pAGN} disc profiles in Figure \ref{fig:C_prof}, showing that the anticipated dissipation $\Delta E_\mathrm{bin}/E_\mathrm{H,bin}$ can vary by 3-4 orders of magnitude due to the density variation across the AGN disc. The increase in $\rho$ and $\mathcal{C}$ for the SG$\beta$ disc model is the result of lower viscosities in the radiation dominated zone, requiring a greater amount of mass at each radius in order to achieve the same Eddington luminosity fraction.

The primary determinant for binary capture is whether the energy condition of Eq.~\eqref{eq:cap_criterion} is satisfied. At separation of $2\hillradius$,
the relative velocity of the two objects is taken to be the combined magnitude of the ambient velocity dispersion in the disc, $\sigma_\mathrm{disp}$ and the Keplerian shear over the radial separation of the BHs $\Delta R$, i.e. $\sigma_\mathrm{Kep}=|R\,\Delta R\, d\Omega/dR|=\frac{3}{2}\Omega \Delta R$. The energy $E_\mathrm{2H}$ is then 
\begin{equation}
    \centering
   E_\mathrm{2H} = \frac{1}{2}\mu v_\mathrm{rel}^{2} -\frac{G \mbin\mu}{2\hillradius}\,.
    \label{eq:E_2H_av_unnorm}
\end{equation}
Using $v_\mathrm{rel}^{2}=\sigma_\mathrm{Kep}^{2}+\sigma_\mathrm{disp}^{2}$, the average encounter energy is
\begin{equation}
    \centering
    E_\mathrm{2H} = \frac{1}{2}\mu (\sigma_\mathrm{Kep}^{2}+ \sigma_\mathrm{disp}^{2}) -\frac{G \mbin\mu}{2\hillradius}\,.
    \label{eq:E2H_midstep}
\end{equation}
Expressing Eq. \eqref{eq:E2H_midstep} in units of $E_\mathrm{H,bin}$ (as required by Eq. \ref{eq:cap_criterion}), this gives 
\begin{equation}
    \centering
   \frac{E_\mathrm{2H}}{|E_\mathrm{H,bin}|} = \frac{\hillradius (\sigma_\mathrm{Kep}^{2}+ \sigma_\mathrm{disp}^{2})}{G\mbin}- 1
   \,.
    \label{eq:E_2H_av0}
\end{equation}
We take the maximum approach velocity to be the shear over, $\Delta R =2\hillradius$, i.e $\sigma_\mathrm{Kep,max} = 3\Omega r_H$. This gives the maximum possible encounter energy as 
\begin{align}
    \centering
   \frac{E_\mathrm{2H}}{|E_\mathrm{H,bin}|}_\mathrm{max} &=
   \frac{\hillradius \Omega^2 (9 \hillradius^2+H^2)}{G\mbin}- 1
   \nonumber =
   9\frac{\mSMBH}{\mbin}\frac{\hillradius^3}{R^3} + \frac{\hillradius H^2}{R^3}- 1
   \\ &=\frac{1}{2}+\left(\frac{\mbin}{6\mSMBH}\right)^{1/3}\frac{H^{2}}{R^2}
   \,.
    \label{eq:E_2H_av}
\end{align}
The ambient velocity dispersion of objects in the disc $\sigma_\mathrm{disp}=Hv_\mathrm{Kep}/R$ is assumed to be equal to the local sound speed, so their mean vertical motion is of order $H$. 
This is a conservative assumption as it assumes BHs that have not yet aligned have zero chance of forming binaries via gas assisted captures and binaries in the disc do not further settle to the midplane. For the assumed parameters, $\sigma_\mathrm{disp}^2/\sigma_\mathrm{Kep}^2$ ranges from $10^{-2}-10^{-4}$.

This formalism does not account for any eccentricity in the BH orbits around the SMBH due to gas effects when embedded objects open up a gap \citep{Sari2004}, additional gravitational focusing on approach to $r=2\hillradius$ or the long-term depletion of gas due to gap opening prior the encounter\footnote{The simulations of \cite{Rowan2023} evolved the BHs for just a few AGN orbits before their encounter.}. Using equations 32 and 33 of \citet{Pan2021} (see also \citealt{Kocsis2011}), the gap opening criteria is only marginally satisfied for $\mSMBH\lesssim 10^{5}\msun$ and $\mSMBH\lesssim 5 \times 10^{5}\msun$ for $\mBH=10\msun$ and $50\msun$ respectively (assuming $\alpha=0.1$). Therefore excluding this effect likely does not affect the predictions of the simulations. Embedded BHs may also have modified eccentricities and or encounter energies from two-body scatterings prior to the encounter. The two body relaxation timescale \citep[calculated according to][]{Tremaine2002,Naoz2022} of the system is $\{0.05,1.5,50\}$Myr for $\mSMBH=\{10^{5},10^{7},10^{9}\}\msun$. This calculation assumes the size of the system is 0.01pc, the RMS BH mass is 10$\msun$ and the confinement of objects to the disc %introduces a factor
increases the density and reduces the two-body relaxation timescale by a factor $\sim(H/0.01\text{pc})^{2}$. Taking the lifetime of the AGN to be $t_\mathrm{AGN}$=10Myr, scattering could be important in the low $\mSMBH$ regime, however note this is an underestimate as it assumes all objects in the system have been embedded in the disc and ignores stars/neutron stars. The resulting additional velocity dispersion from two-body scatterings could potentially reduce the merger rate via an increase in the value of $\sigma_\mathrm{disp}$ in Eq. \eqref{eq:E_2H_av0}--\eqref{eq:E_2H_av}.

\begin{figure}
    \centering
    \includegraphics[width=8cm]{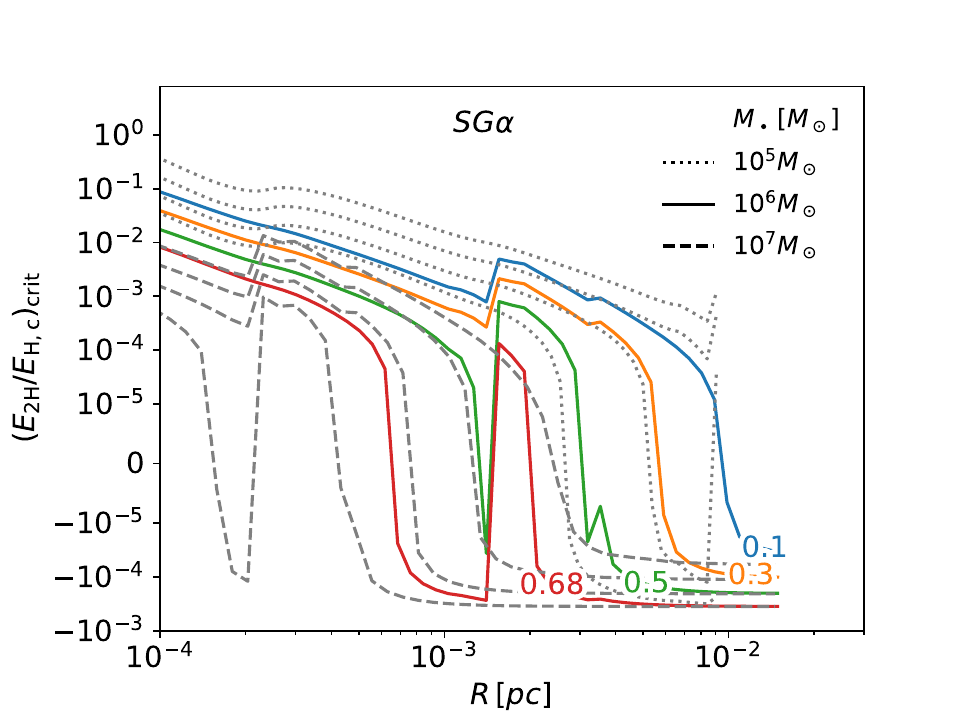}
    \caption{Maximum initial encounter energy $E_\mathrm{2H}$ of a binary that leads to a successfully formed binary for different impact parameters $\ph$, as labelled on the curves, as a function of radial distance in the AGN disc $R$. Results shown for an SG$\alpha$ disc with parameters $\mbin=20\msun$ and  $\mSMBH=[10^{5},10^{6},10^{7}]\msun$. At lower $R$, the lower velocity dispersion and higher gas density allows BHs with larger initial encounter energies to dissipate enough energy to stay bound. Closer encounters at low impact parameters can extend binary formation to larger $R$.}
    \label{fig:E_crit_vs_R}
\end{figure}

In Figure \ref{fig:E_crit_vs_R} we show the maximum allowable encounter energies $(E_\mathrm{2H}/E_\mathrm{crit})$ that lead to successful binary formation for select values of $p_\mathrm{1H}$, assuming three values of $\mSMBH$.
The contours of $(E_\mathrm{2H}/E_\mathrm{crit})$ with $p_\mathrm{1H}$ at low $R$ form a steep power law where initially unbound BHs can still be captured into binaries, driven by the first term in equation \eqref{eq:cap_criterion_full}. Then, as $\mathrm{C}$ decreases at higher $R$, the second term in Eq. \eqref{eq:cap_criterion_full} dominates and only encounters with increasingly negative energies will form stable binaries, tending towards an asymptotic value dependent on $\ph$. Decreasing $p_\mathrm{1H}$ values lead to higher dissipation values and allows binaries to be more easily formed at higher energies further out in the AGN disc (for the same typical encounter energy). The transition point from the power law to the plateau occurs at lower $R$ for smaller $\mSMBH$. In the low $\mSMBH$ regime, the density profile of the AGN exhibits an additional minimum due to the opacity break in the inner region of the disc, leading to a region of inefficient binary formation, suggesting a region where the density of gas-captured binaries could be lower than anticipated otherwise. However, the number of BHs in this region, particularly for low $\mSMBH$, is likely not significant. 
\subsection{The formation function}
\begin{figure}
    \centering
    \includegraphics[width=9cm]{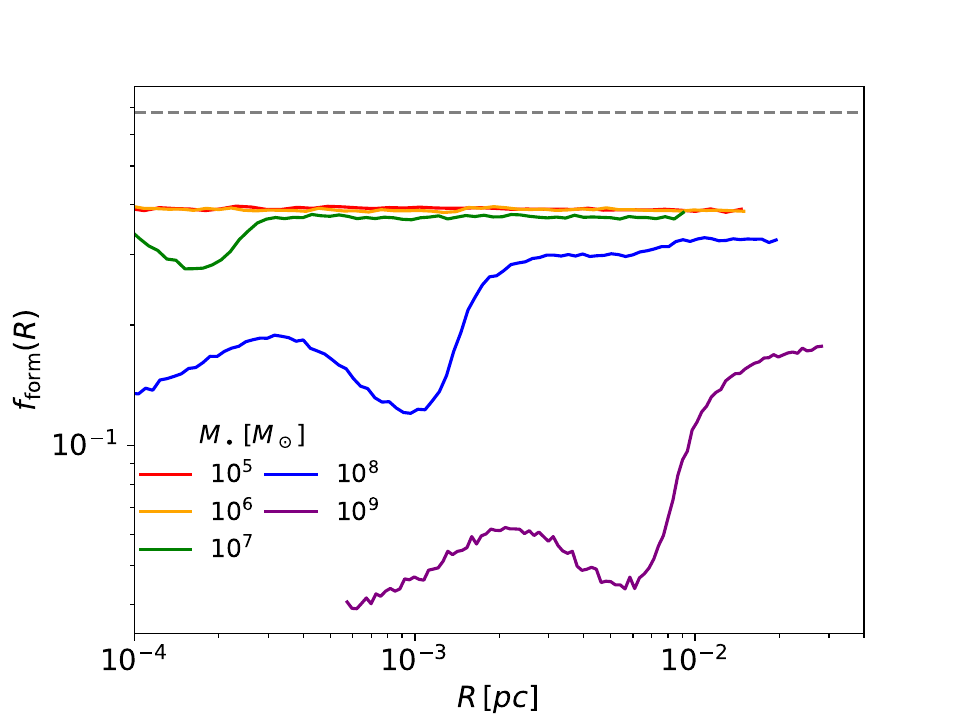}
    \caption{Fraction of encounters with impact parameters $p_\mathrm{1H}<\hillradius$ that lead to successfully formed binaries as a function of radial distance in the SMBH disc for different $\mSMBH$, assuming an SG$\alpha$ disc. The function is shown assuming the BH mass function of BIMF$^{\gamma=2.35}_\text{Tagawa}$ and a uniform distribution of $p_\mathrm{1H}$ and $\Delta R$ is assumed. Results show AGN with higher $\mSMBH$ have a reduced formation probability for higher $R$.}
    \label{fig:f_form_R_monte_carlo}
\end{figure}

To determine the rate of BH mergers, it is necessary to define the statistical fraction of BH encounters that will lead to binary formation $f_\mathrm{form}(R,M_\mathrm{bin},\mSMBH)$. Here, encounters are defined as events where the separation of two BHs is less than the binary Hill radius. In Figure \ref{fig:f_form_R_monte_carlo}, $f_\mathrm{form}$ is shown as a function of $R$ for a range of $\mSMBH$, sampling the initial mass distribution from BIMF$_\mathrm{Tagawa}$. 
The ambient velocity dispersion term $\sigma_\mathrm{disp}$ is sampled from a random Gaussian distribution with standard deviation $\sigma\equiv c_\mathrm{s}$ centred on zero. The upper bound of $\Delta R$ in accordance with Eq. \eqref{eq:E_2H_av} is taken to be $2\hillradius$. This is also a conservative assumption as \citet{Rowan2022},\citet{Whitehead2023} and \citet{Rowan2023} indicated BHs can be focused into minimum encounter separations smaller than $\hillradius$ for $\Delta R>2\hillradius$ depending on the disc density. The values of $\Delta R$ and $p_\mathrm{1H}$ are sampled uniformly\footnote{In practice there is a correlation between $p_\mathrm{1H}$ and $\Delta R$, however as it is unclear how this should be affected by the ambient velocity dispersion $\sigma_\mathrm{disp}$, they are sampled randomly.} between $0.01\hillradius-2\hillradius$ and $0-\hillradius$ respectively for approaching BHs, where the lower $\Delta R$ value is given a non-zero value to avoid divergence in the calculation of the time between encounters later in Sec. \ref{sec:timescales}.

The formation function is weakly dependent on $\mSMBH$, with a decrease of roughly an order of magnitude in the formation probability from $\mSMBH=10^{5}\msun$ to $10^{9}\msun$. The formation probability becomes flat for $\mSMBH<10^{7}\msun$ as the gas dissipation becomes highly efficient at forming binaries from typically lower relative velocities ($E_{\rm 2H}/E_{\rm H,c}\propto\mSMBH^{2/3}$). Similarly, the formation probability is larger at higher radii in the AGN disc. The initial increases in $\fform$ result from the  maxima in the density ($\rho$) profiles, which shifts to higher $R$ for larger $\mSMBH$. 
\subsection{BH binary merger rates using a Monte Carlo approach}
\label{sec:monte_carlo}
In order for an isolated black hole to merge, it must satisfy four conditions within $t_\mathrm{AGN}$: \textit{i}) the BHs must align with the disc, \textit{ii}) encounter another BH, \textit{iii}) successfully form a binary and \textit{iv}) successfully merge. To estimate the overall merger rate, we first consider the timescale for a single BH to go through each of these stages in its evolution. 
\subsubsection{The timescales of the system}
\label{sec:timescales}
We assume the alignment timescale is derived according to \citet{Bartos2017}. Given some initial vertical velocity $v_z$ at the point of disc crossing and a typical velocity reduction $\Delta v_z$ upon crossing the disc from dynamical friction, the general expression for the characteristic timescale of disc alignment is
\begin{equation}
    \centering
    t_\mathrm{align}\simeq\frac{t_\mathrm{orb}}{2}\frac{v_z}{\Delta v_z}\,.
    \label{eq:t_align_fulla}
\end{equation}
The $t_\mathrm{orb}=2\pi R^{3/2}(G\mSMBH)^{-1/2}$ term is the orbital period, and the factor 2 accounts for two crossings per $t_\mathrm{orb}$. The fractional change in velocity of an object is equated to the ratio of the mass accreted during its crossing of the disc and its own mass such that $\Delta v_z/v_z=\Delta M_\mathrm{cross}/\mBH$. The accreted mass is assumed to be that within its Bondi-Hoyle-Lyttleton radius $r_\mathrm{BHL}=2G\mBH/(\Delta v^{2}+c_s^{2})$, where $\Delta v=v_\mathrm{orb}((1-\cos(i))^{2}+\sin^{2}(i))^{1/2}=2v_\mathrm{orb}\sin(\frac{i}{2})$ is the relative velocity of the binary to the gas, which orbits the SMBH with velocity $v_\mathrm{orb}=\sqrt{G\mSMBH/R}$. The crossing mass is then
\footnote{Note that in reality $r_{\rm BHL}$ should be replaced by $\min(r_{\rm BHL},\hillradius)$, but $r_{\rm BHL}/\hillradius< \frac{3^{1/3}}{4 \sin^2(i/2)}(\mBH/\mSMBH)^{2/3}\ll 1$ for $i \gg 3^{1/6}(\mBH/\mSMBH)^{1/3}$. The final alignment may be  prolonged if $H/R < 3^{1/6}(\mBH/\mSMBH)^{1/3}$. Note that near the disk's self-gravitating boundary $H/R\sim M_{\rm d}/2\mSMBH$ (see Eq.~\ref{eq:Toomre} and $c_{\rm s}=\Omega H$), implying a prolonged relaxation if $M_{\rm d}< (\mBH/\mSMBH)^{1/3} \mSMBH$.} $\Delta M_\mathrm{cross}=\Delta v t_\mathrm{cross}r_\mathrm{BHL}^{2}\pi\Sigma/(2H)$ with crossing time $t_\mathrm{cross}\approx2H/(v_\mathrm{orb}\sin{i})$. 
Putting all this together gives an alignment time of 
\begin{align}
    t_\mathrm{align}&=\frac{t_\mathrm{orb}}{2}\frac{\cos(i/2)(\Delta v^{2}+c_s^{2})^{2}}{4G^{2}\mBH\pi\Sigma}\nonumber\\
    &=\frac{ t_{\rm orb} \mSMBH^2}{2 M_{\rm BH}M_{\rm d}}
     \cos\left(\frac{i}2\right)\left[\sin^2\left(\frac{i}2\right)+\frac{H^2}{4R^2}\right]^{2}\,,    
    \label{eq:t_align_full}
\end{align}
where the identity $\sin(i)/\sin(i/2)=2\cos(i/2)$ has been applied and we define $M_{\rm d}=2\pi R^2\Sigma$. 
Note that the second term in the parenthesis $H^2/4R^2$ is much smaller unless the orbit is close to being fully embedded in the disk where $\sin (i/2) \approx \frac12 \sin i = H/2R$.
Note the strong dependence on the velocity term $(\Delta v^{2}+c_s^{2})^{2}$ to the fourth power, which makes it increasingly difficult to embed objects for larger SMBH masses for a fixed $R$ or $M_{\rm d}$. Note that the potential of the AGN disc or stellar population is not accounted for.

The encounter timescale $t_\mathrm{enc}$ is given by:
\begin{equation}
    t_\mathrm{enc}=\frac{1}{n_\mathrm{BH}\hillradius z_\mathrm{H}v_\mathrm{rel}}
    \approx
    \frac{2/3}{n_\mathrm{BH}\Omega\hillradius^3} = 
    \frac{4}{n_\mathrm{BH}G^{1/2}\mSMBH^{-1/2}\mbin R^{3/2}}
    \,,
    \centering
    \label{eq:t_enc}
\end{equation}
where $n_\mathrm{BH}$ is the volume number density of BHs, $z_\mathrm{H}=\min(H,\hillradius)$ is the vertical cross section for the encounter in the case $\hillradius<H$.\footnote{Similar to the conditions required to open a gap in the disc, the vertical cross section $z_\mathrm{H}$ is only less than $\hillradius$ for low $\mSMBH$, here typically $\mSMBH<5\times10^{5}\msun$.} We assume the relative velocity is equivalent to the velocity shear across the entire Hill sphere $v_\mathrm{rel}=\frac{3}{2}\Omega\,\hillradius$ (since objects can approach from inside and outside the BHs orbit). In practice $v_\mathrm{rel}$ is likely larger due to the assumed velocity dispersion of BHs and the ability for an approaching BH to be perturbed from higher/lower radii to a radial separation of $\hillradius$ due to gravitational focusing \citep[e.g][]{Boekholt_2022}. Once a BH encounters another in the disc, the formation likelihood is given by $\fform$. This then modifies $t_\mathrm{enc}$ to give the effective formation timescale 
\begin{equation}
    t_\mathrm{form}=\frac{t_\mathrm{enc}}{\fform}\,.
    \centering
    \label{eq:t_enc_eff}
\end{equation}
Determining the value of $\fform$ requires knowledge of the expected number and masses of BHs at a particular point of the disc, which we discuss in Sec.~\ref{sec:formation_probability} below.

Perhaps the most uncertain timescale is the merger timescale. It has been shown that while retrograde binaries can reliably merge \citep[e.g.][]{Li_and_lai_2022,Li_and_Lai_2022_windtunnel_II,Rowan2022,Li_and_Lai_windtunnel_III_2023}, prograde binaries can in some cases outspiral \citep[e.g.][]{Li2021,Dempsey2022}. These binaries are typically given zero initial eccentricity, though it has been shown that eccentricity persists long after the initial formation. Additionally it has been shown that hotter circum-binary discs (more typical when the isothermal assumption is relaxed) lead to reliable inspiral \citep[e.g.][]{Baruteau2011,Li_2022_hot_discs}. In this work, like most other population studies \citep[e.g.][]{Tagawa2020,Mapelli2021,ford2022_AGNrates}, it is assumed that the binaries formed are reliably hardened by gas. The merger timescale itself is significantly smaller than the AGN lifetime \citep[e.g.][]{Haehnelt1993,Cavaliere1989}. Nevertheless, given the uncertainty of the inspiral rate still present in the literature, the maximal merger timescale in \citet{Bartos2017} of $t_\mathrm{merge}\sim10^{5}$yr is used corresponding to $\mSMBH=10^{6}\msun$ at 0.01pc. At higher $\mSMBH$ or lower $R$, the inspiral rate is shorter, but the value is maintained for all binaries as pessimistic assumption. Even at this upper bound, the merger timescale is still two orders of magnitude shorter than the AGN lifetime, leaving $t_\mathrm{align}$ and $t_\mathrm{enc}$ as the more impactful timescales for calculating merger rates.

The full timescale of a BH to merge, including the dependencies is
\begin{equation}
    t_\mathrm{tot}=t_\mathrm{align}(R,\mBH)+t_\mathrm{form}(R,\mBH,n_\mathrm{BH})+t_\mathrm{merge}\,.
    \centering
    \label{eq:total_lifetime}
\end{equation}
The merger rate is then the number of BHs which satisfy $t_\mathrm{tot}<t_\mathrm{AGN}$ divided by $2t_\mathrm{AGN}$, where the factor 2 accounts for double counting. 
\subsection{Resolving dependencies}
\subsubsection{BH number density}
\begin{figure}
    \centering
    \includegraphics[width=8cm]{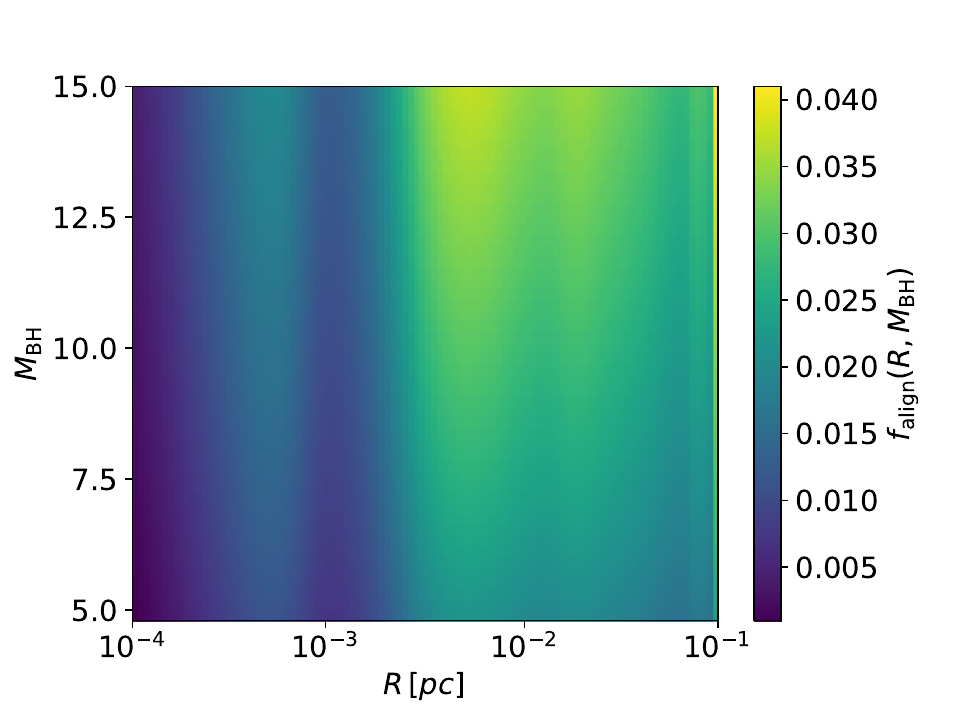}
    \caption{Fraction $f_\mathrm{align}$ of BHs with mass $\mBH$ at radius $R$ aligning with an SG$\alpha$ AGN disc for fiducial parameters $\mSMBH=4\times10^{6}\msun$, BIMF$_\text{Tagawa}^{\gamma=2.35}$ and $t_\mathrm{AGN}=10^{7}$yr. Generated by evaluating equation Eq. \eqref{eq:t_align_full} over uniform $\cos{i}$. Figure shows higher mass BHs can align quicker, with a larger fraction aligning typically at larger radii.}
    \label{fig:p_align}
\end{figure}

The complexities of the calculation lie within the dependencies of $n_\mathrm{BH}$ and $\fform$. Starting with the former, the number density of BHs in the disc $n_\mathrm{BH}$ is determined from the initial sample of $\{M_{\mathrm{BH},i}\}$ and $\{R_i\}$ (see Sec. \ref{sec:pop_sample}) that satisfy $t_\mathrm{align}<t_\mathrm{AGN}$. To smooth out the stochasticity in the sampling from the calculation of $n_\mathrm{BH}$, a probability grid, $f_\mathrm{align}(R,\mbin)$, in bins of $\mBH$ and $R$ is constructed by sampling over the full range of $M_{\mathrm{BH},i}$ and $R_i$,  calculating their alignment time using Eq.~\eqref{eq:t_align_full}.  For each bin in $\mBH$ and $R$, the alignment timescale is sampled over the full range of $\cos{i}$ and the probability of that BH to align is the number of instances where the alignment time condition is met as a fraction of the number of inclination samples. The grid is constructed in 100 bins of $R$ in log space and $50$ bins of $\mBH$ in linear space. For a fiducial setup of $\mSMBH=4\times10^{6}\msun$, the probability of alignment across $\mBH$ and $R$ is shown in Figure \ref{fig:p_align}. As predicted by Eq. \eqref{eq:t_align_full}, the figure indicates BHs at smaller $R$ embed themselves more easily in the AGN disc.
\tikzset{%
  >={Latex[width=2mm,length=2mm]},
  % Specifications for style of nodes:
            base/.style = {rectangle, rounded corners, draw=black,
                           minimum width=4cm, minimum height=1cm,
                           text centered},
            logic_base/.style = {rectangle, rounded corners, draw=black,
                           minimum width=0.5cm, minimum height=0.4cm,
                           text centred},
  activityStarts/.style = {base, fill=blue!30},
       startstop/.style = {base, fill=red!30},
    activityRuns/.style = {base, fill=green!30},
         process/.style = {base, minimum width=2.5cm, fill=orange!15},
         logic/.style = {logic_base, minimum width=0.5cm, fill=orange!15},
}
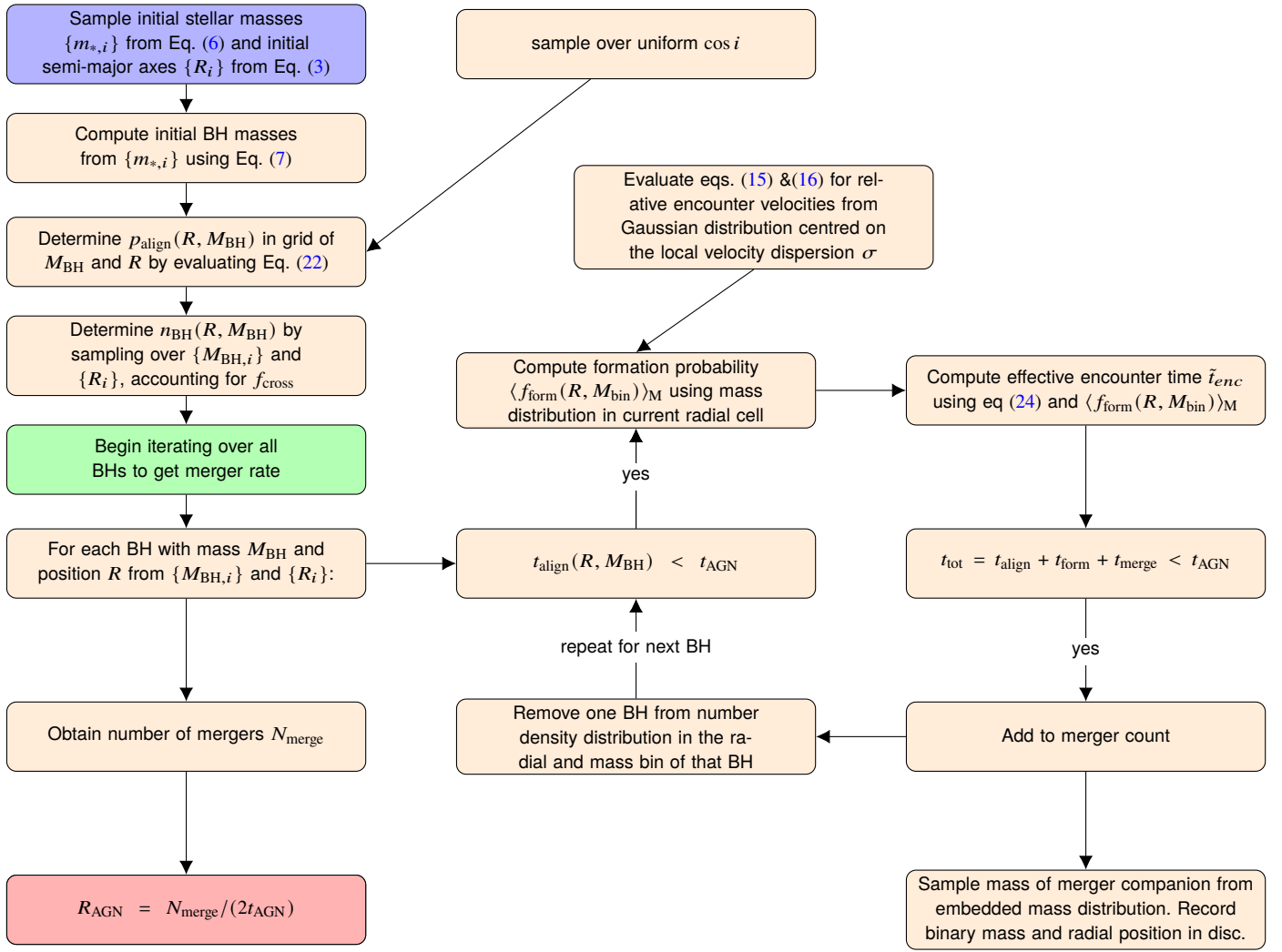
\begin{figure*}
\begin{tikzpicture}[node distance=1.5cm,
    every node/.style={fill=white, font=\sffamily,text width=5.cm}, align=center]
  % Specification of nodes (position, etc.)
  \node (start)             [activityStarts]              {Sample initial stellar masses $\{m_{*,i}\}$ from Eq. \eqref{eq:kroupa} and initial semi-major axes $\{R_i\}$ from Eq. \eqref{eq:Radial_dist}};
  \node (onCreateBlock)     [process, below of=start]          {Compute initial BH masses from  $\{m_{*,i}\}$ using Eq. \eqref{eq:BIMF_tagawa}};
  \node (onStartBlock)      [process, below of=onCreateBlock]   {Determine $p_\mathrm{align}(R,\mBH)$ in grid of  $M_{\mathrm{BH}}$ and $R$ by evaluating Eq. \eqref{eq:t_align_full}};
  \node (onResumeBlock)     [process, below of=onStartBlock]   {Determine $n_\mathrm{BH}(R,\mBH)$ by sampling over $\{M_{\mathrm{BH},i}\}$ and $\{R_i\}$, accounting for $f_\mathrm{cross}$};
  \node (activityRuns)      [activityRuns, below of=onResumeBlock]
                                                      {Begin iterating over all BHs to get merger rate};
  \node (getBH_i)           [process,below of=activityRuns] {For each BH with mass $\mBH$ and position $R$ from $\{M_{\mathrm{BH},i}\}$ and $\{R_i\}$:};      

  \node (align_cond)      [process, right of=getBH_i, xshift =5cm] {$t_\mathrm{align}(R,\mBH)<t_\mathrm{AGN}$};

  \node (compute_f_form)      [process, above of=align_cond, yshift = 1cm] {Compute formation probability $\langle \fform(R,M_\mathrm{bin})\rangle_\mathrm{M}$ using mass distribution in current radial cell};

\node (compute_f_form_detail)      [process, above of=compute_f_form, yshift = 1cm,xshift = 1.7cm] {Evaluate eqs. \eqref{eq:cap_criterion} \&\eqref{eq:cap_criterion_full} for relative encounter velocities from Gaussian distribution centred on the local velocity dispersion $\sigma$};

  \node (compute_t_enc)      [process, right of=compute_f_form, xshift = 5cm] {Compute effective encounter time $\Tilde{t}_{enc}$ using eq \eqref{eq:t_enc_eff} and $\langle \fform(R,M_\mathrm{bin})\rangle_\mathrm{M}$};

  \node (merge_check)      [process, below of=compute_t_enc, yshift = -1cm] {$t_\mathrm{tot}=t_\mathrm{align}+t_\mathrm{form}+t_\mathrm{merge}<t_\mathrm{AGN}$};

  \node (sample_merg_props)      [process, below of=merge_check, yshift = -1cm] {Add to merger count};

  \node (record_merg_props)      [process, below of=sample_merg_props, yshift = -1cm] {Sample mass of merger companion from embedded mass distribution. Record binary mass and radial position in disc.};

  \node (get_Nmerg)      [process, below of=getBH_i, yshift=-1.cm] {Obtain number of mergers $N_\mathrm{merge}$};

  \node (get_rate)      [startstop, below of=getBH_i, yshift=-3.5cm] {$R_\mathrm{AGN}=N_\mathrm{merge}/(2t_\mathrm{AGN})$};

  \node (update_n_disc)      [process, left of=sample_merg_props, xshift = -5cm] {Remove one BH from number density distribution in the radial and mass bin of that BH};
                                                      
  \node (get_rate)      [startstop, below of=getBH_i, yshift=-3.5cm] {$R_\mathrm{AGN}=N_\mathrm{merge}/(2t_\mathrm{AGN})$};

  \node (onRestartBlock)    [process, right of=onStartBlock, xshift=5cm,yshift = 3cm]
                                                              {sample over uniform $\cos{i}$};
    
  % Specification of lines between nodes specified above
  % with aditional nodes for description 
  \draw[->]             (start) -- (onCreateBlock);
  \draw[->]     (onCreateBlock) -- (onStartBlock);
  \draw[->]      (onStartBlock) -- (onResumeBlock);
  \draw[->]     (onResumeBlock) -- (activityRuns);
  \draw[->]     (activityRuns) -- (getBH_i);

  \draw[->]    (onRestartBlock) -- (onStartBlock.east);
  
  \draw[->] (align_cond) -- (compute_f_form);
  \draw[->] (getBH_i) -- (align_cond);

  \draw[->] (compute_f_form) -- (compute_t_enc);

  \draw[->] (compute_t_enc) -- (merge_check);

  \draw[->] (merge_check.south) --               
     node[yshift=-0.5, text width=2.5cm]
     {yes}(sample_merg_props.north);

 \draw[->] (sample_merg_props) -- (record_merg_props);

 \draw[->] (sample_merg_props) -- (update_n_disc);

  \draw[->] (update_n_disc.north) --               
     node[yshift=0.5, text width=2.5cm]
     {repeat for next BH}(align_cond.south);

  \draw[->]    (compute_f_form_detail.south) -- (compute_f_form.north);

  \draw[->]    (getBH_i) -- (get_Nmerg);

  \draw[->]    (get_Nmerg) -- (get_rate);

  \draw[->] (align_cond.north) --               
     node[yshift=0.5, text width=2.5cm]
     {yes}(compute_f_form.south);

%  \draw[->] (getBH_i.east) -- ++(2.6,0) -- ++(0,2) -- ++(0,2) --                
%     node[xshift=1.2cm,yshift=-1.5cm, text width=2.5cm]
%     {The activity comes to the foreground}(onResumeBlock.east);
  \end{tikzpicture}
\centering
\caption{Summary of the fiducial semi-analytic procedure to determine black hole merger rates, from an initial sample of stars in the central stellar cluster.}
\label{fig:flowchart}
\end{figure*}
The number density is similarly calculated as a grid in $\mBH$ and $R$ by sampling over the full initial distributions of $\{M_{\mathrm{BH},i}\}$ and $\{R_i\}$, binning them into the same bins for $p_\mathrm{align}$ and then adding the probability for that BH to align with the disc. The number density is also represented as a grid in $\mBH$ and $R$ to keep track of the mass distribution at each radius, as this affects the calculation of $\fform$. When evaluating $t_\mathrm{enc}$ in Eq. \eqref{eq:t_enc}, the number density is the sum of the number densities across the mass bins.
\subsubsection{Formation probability}
\label{sec:formation_probability}
The BH formation function $\fform$ is a function of both $R$ according Eq. \eqref{eq:cap_criterion_full} but also the anticipated binary mass by Eq. \eqref{eq:E_2H_av}. For a given BH we compute the mass averaged value of the formation function $\langle \fform \rangle_\mathrm{M}$ by sampling over the mass distribution at the radial position of the BH, utilising the mass distribution of $n_\mathrm{BH}(R,\mBH)$. Since smaller objects are less likely to align in the AGN disc, the mass distribution of embedded BHs skews to higher masses compared to the initial distribution.
\subsubsection{Comparing to dynamical friction}
We will compare our results to another suite of Monte Carlo simulations where $\fform$ is determined assuming a simplified gas dynamical friction treatment. There, we incorporate the \citet{Tagawa2020} formation prescription into the Monte Carlo simulations in the calculation of the formation timescale. Under this prescription, the probability that a BH-BH scattering successfully forms a binary is given by 
\begin{equation}
    \centering
    f_{\rm form,Tag}=\min(1,t_{\rm pass}/t_{\rm DF})\,,
    \label{eq:fform_tag}
\end{equation}
where $t_{\rm pass}$ is the time taken for the objects to traverse the Hill sphere,
\begin{equation}
    \centering
    t_{\rm pass}=\hillradius/ v_{\rm rel}\,,
    \label{eq:t_pass}
\end{equation}
and $t_{\rm DF}=v_{\rm rel}/a_{\rm DF}$ is the gas dynamical friction timescale from the deceleration given by \citep{Ostriker1999}:
\begin{align}
    \centering
    a_{\rm DF}(v_\mathrm{rel})&=-\frac{4\pi G^{2}\mBH\rho}{v_{\rm rel}^{2}}f(v_{\rm rel}/c_\mathrm{s})\,,\\
    f(x) &= \begin{cases}
      \frac{1}{2} \ln \big(\frac{1+x}{1-x}\big)-x  & 0<x\leq1\,, \\
      \frac{1}{2}\ln{(x^{2}-1)}+3.1  & x>1\,.
    \end{cases} 
    \label{eq:a_DF}
\end{align}
The modified formation function in Eq. \eqref{eq:fform_tag} is used in Eq. \eqref{eq:t_enc_eff} in the same manner as before, with identical sampling of the relative velocity (Sec. \ref{sec:monte_carlo}).
\subsection{Knock on effects}
Extending the single BH calculation to a sample across the entire BH population requires accounting for knock on effects from the outcome of each calculation, i.e. was there a merger. To account for the finite number of BHs and the time dependence of $n_\mathrm{BH}$ the contribution of one BH is removed from the number density for each merger. Specifically, if a BH satisfies the time constraints of Eq. \eqref{eq:total_lifetime}, a random BH merging partner is sampled from the current distribution of masses predicted by $n_\mathrm{BH}$ and its contribution to the number density is removed. The masses and position in the disc is recorded for all mergers, which is required to compute the anticipated merger rate from GW detectors. Though this is not a formal implementation of the time dependence of $n_\mathrm{BH}$ and one cannot comment on the change in merger rate over the AGN lifetime, it accounts for the overall reduction in the number of mergers within $t_\mathrm{AGN}$. Additionally, neglecting this effect would lead to over-counting higher mass binary encounters/mergers, since higher mass BHs have shorter encounter timescales. From Eq. \eqref{eq:t_enc} there is an overall $\mbin$ dependence of $\sim\mbin^{-1}$ (assuming $z_\mathrm{H}=\hillradius$, which is true for the vast majority of $R,\mbin$ and $\mSMBH$).  

However, the merger rate dependence on this assumption is small, as discussed in Sec. \ref{sec:observable_rates}. For clarity, a detailed flow chart to summarise the overall Monte-Carlo Scheme is shown in Figure \ref{fig:flowchart}.
\section{Results}
\label{sec:results}
\subsection{$\mSMBH$ dependence}
\label{sec:SMBH_dependence}
\begin{figure}
    \centering
    \includegraphics[width=7cm]{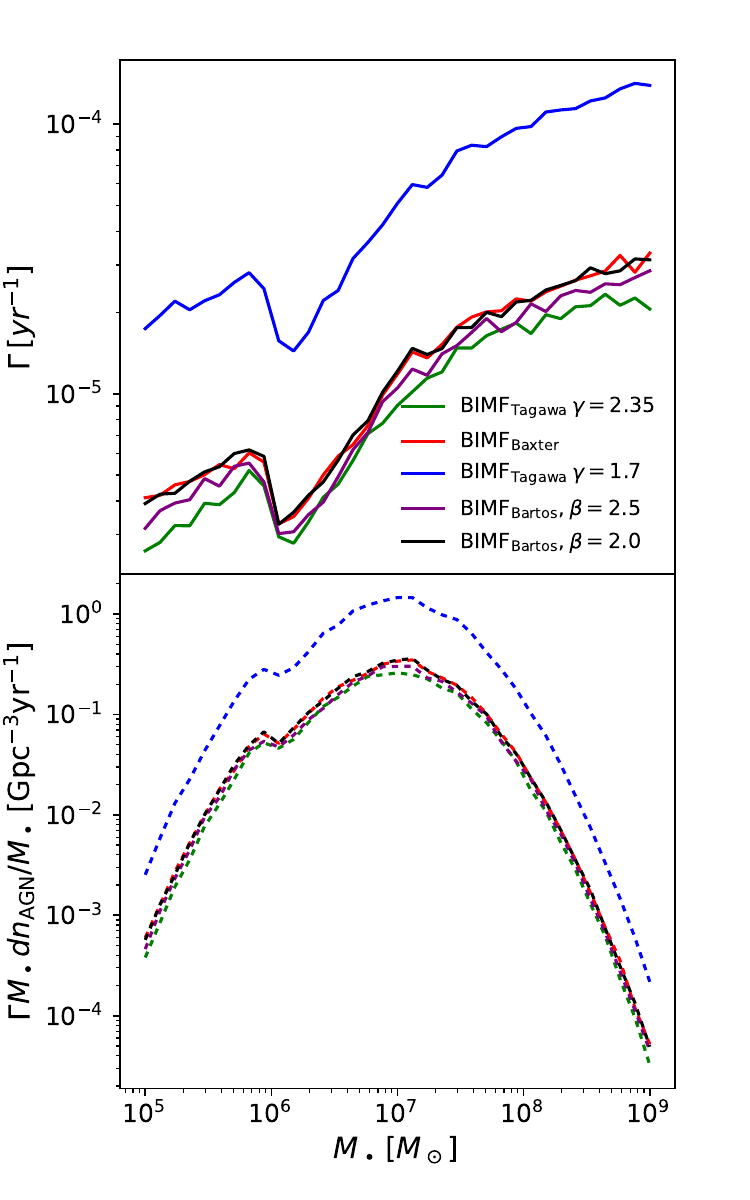}
    \caption{\textit{Top:} The BH merger rate $\Gamma$ per year per AGN with mass $\mSMBH$ for the SG$\alpha$ models. \textit{Bottom:} The resulting merger rate across $\mSMBH$ weighted by the mass distribution of AGN (Eq. \ref{eq:AGN_mass_funct}). The different colours represent the assumed BH initial mass function (see Sec. \ref{sec:pop_sample}). The graph indicates the observable rates should be dominated by AGN with $\mSMBH\sim 10^{7}\msun$.}
    \label{fig:rates_over_M_SMBH_fiducial}
\end{figure}

The merger rate \textit{per} AGN per year ($\Gamma$) and observable merger rate density distribution (considering only the SG$\alpha$ models for now) over $\mSMBH$ is shown in Figure \ref{fig:rates_over_M_SMBH_fiducial}. The merger rate distribution assumes the AGN number density in the Universe follows the function given by \citet{Greene_Ho2007,Greene2009}
\begin{equation}
\frac{dn_\mathrm{AGN}}{d\mSMBH}=\frac{34000\text{ Gpc}^{-3}}{\mSMBH}10^{-(\log_{10}(\mSMBH/M_\odot)-0.67)^{2}/1.22}\,.
    \centering
    \label{eq:AGN_mass_funct}
\end{equation}
The rate of BH mergers is dependent on SMBH mass, ranging from order $\sim10^{-6}$yr$^{-1}$ at $\mSMBH=10^{5}\msun$ to $\sim10^{-4}$yr$^{-1}$ at $\mSMBH=10^{9}\msun$ for a single AGN. 
\begin{figure*}
    \centering
    \includegraphics[width=16cm]{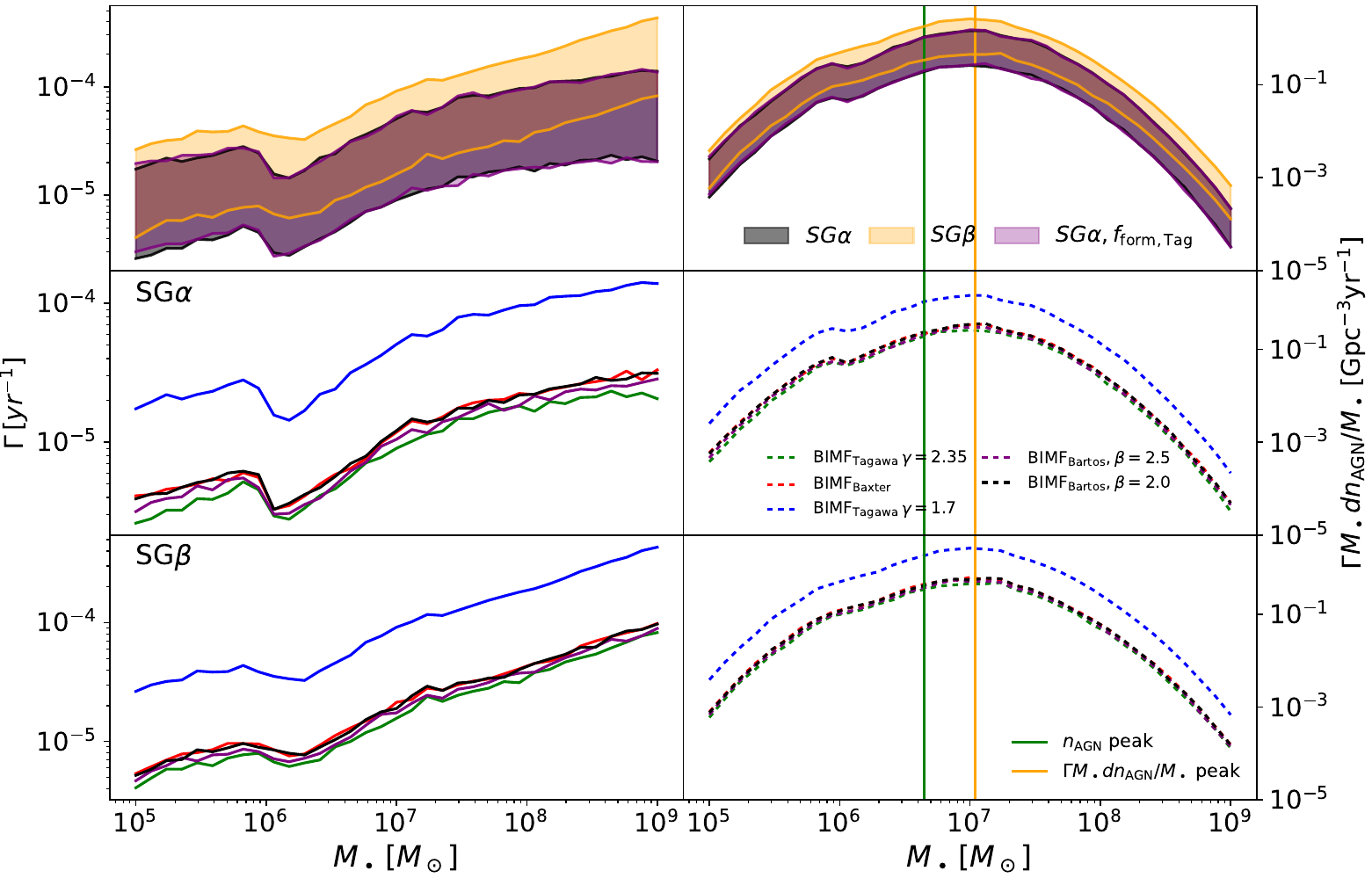}
    \caption{\textit{Left column:} The BH merger rate $\Gamma$ per year per AGN with mass $\mSMBH$. \textit{Right column:}  The merger rate across $\mSMBH$ weighted by the mass distribution of AGN (Eq. \eqref{eq:AGN_mass_funct}). The vertical green and yellow lines indicate the peak of the AGN mass function and the rate distribution over $\mSMBH$ respectively. 
    \textit{Top row:} The range in rates across all BIMFs for each disc viscosity prescription (SG$\alpha$, SG$\beta$) and simplified formation function ($f_\mathrm{form,Tag}$). \textit{Middle row:} the rates for the $SG\alpha$ disc model. \textit{Bottom row:} the rates for the $SG\beta$ disc model.
    The graph indicates the observable rates should be dominated by AGN with $\mSMBH\sim10^{7}\msun$. The SG$\beta$ discs show a greater merger rate, particularly in the high $\mSMBH$ range.}
    \label{fig:rates_over_M_SMBH}
\end{figure*}
These results are in agreement with the lower range predicted by \citet{Tagawa2020} ($\Gamma\sim10^{-4}-10^{-3}$yr$^{-1}$, for $\mSMBH=10^{7}\msun$). The rates are around 10-50 times larger than the results of \citet{Bartos2017}. As \citet{Bartos2017} considers only pre-existing binary mergers (i.e., the binaries did not form inside the AGN disc), the steeper dependence on $\mSMBH$ here is a result of the binary formation function and encounter timescale, which have additional $\mSMBH$ dependence. At low masses, merger rates are restricted by the number of BHs in the system, the lowest number being $\sim100$ BHs for $\mSMBH=10^{5}\msun$. As $\mSMBH$ increases, the number of available BHs within $R_\mathrm{inf}$ increases, at a faster rate than $f_\mathrm{cross}$ can limit the embedded number of BHs. The increased BH population in combination with the AGN mass function results in a peak in the merger rate at $\mSMBH\sim10^{7}\msun$. Beyond this, the scarcity of AGN with $\mSMBH>10^{7}\msun$ limits the contribution to the merger rate despite $\Gamma(\mSMBH)$ being larger for these more massive AGN. 

To investigate the main bottleneck of the merger process, we can consider the fraction of BHs that have $t_\mathrm{align}<t_\mathrm{AGN}$ (F1), then the fraction of F1 which satisfy $t_\mathrm{align}$ + $t_\mathrm{form}<t_\mathrm{AGN}$ (F2) and the fraction of F2 that satisfy $t_\mathrm{align}$ + $t_\mathrm{form}$ + $t_\mathrm{merge}<t_\mathrm{AGN}$ (F3).  Assuming BIMF$_\mathrm{Tagawa}^{\gamma=2.35}$, these values are shown in Table \ref{tab:bottleneck} for $\mSMBH=\{10^{5},10^{7},10^{9}\}\msun$.
\begin{table}
\begin{tabular}{|c|c|c|c|} 

 \hline
     $\mSMBH\,[\msun]$       & F1(\%) & F2(\%) & F3(\%) \\
 \hline
 $10^{5}$  & 26 & 43 & $>99$ \\ 
 \hline
 $10^{7}$ & 1.4  & 65 & $>99$ \\
 \hline
 $10^{9}$ & 0.21  & 80 & $>99$  \\
 \hline
\end{tabular}
\centering
\caption{The fraction of BHs in the Monte Carlo SG$\alpha$ simulation with BIMF$_\mathrm{Tagawa}^{\gamma=2.35}$ that satisfy $t_\mathrm{align}<t_\mathrm{AGN}$ (F1) and of those $t_\mathrm{align}$ + $t_\mathrm{form}<t_\mathrm{AGN}$ (F2) and of those the fraction with $t_\mathrm{align}$ + $t_\mathrm{enc}$ + $t_\mathrm{form}<t_\mathrm{AGN}$ (F3). Indicating that the majority of BHs do not merge due to the aligment time.}
\label{tab:bottleneck}
\end{table}
The relative fractions imply that (per BH) the primary bottleneck is the alignment time (F1) by approximately an order of magnitude, therefore constraining the inclination and radial distribution of BHs in the initial distribution is also crucial. Though F1 is the primary bottleneck, the formation timescale (unlike the merger timescale) is not negligibly small as a notable fraction of embedded BHs fail to form a binary within $t_\mathrm{AGN}$. 

There is no significant (order of magnitude) difference in the overall merger rates from our fiducial model with BIMF$_\text{Baxter}$ and BIMF$_\text{Bartos}$ (Table \ref{tab:merger_rates}). However we find a strong dependence on $\gamma$ for BIMF$_\text{Tagawa}$, with the lower value of $\gamma=1.7$ leading to a rate increase of about an order of magnitude. This stems from having many BHs from the more top heavy stellar mass distribution ($\sim$5 times more BHs) and the resulting top heavy BIMF compared with BIMF$_\mathrm{Tagawa}^{\Gamma=2.35}$. The increase in merger rate is driven by a higher embedded BH density $n_\mathrm{BH}$ since higher mass BHs more easily embed themselves within $t_\mathrm{AGN}$ and have shorter encounter timescales $t_\mathrm{enc}\propto n_\mathrm{BH}^{-1}\mbin^{-1}$. As the range in $\mBH$ ($\mbin$) for BIMF$_\mathrm{Tagawa}$ is only $5\msun-15\msun$ ($10\msun-30\msun)$, this suggests the increase in overall BH number is the dominant factor in the increased rates for $\gamma=1.7$. Allowing for larger initial BH masses increases the rates slightly (up to a factor of a few, e.g BIMF$_\mathrm{Baxter}$) compared to BIMF$^{\gamma=2.35}_\mathrm{Tagawa}$. This comparison highlights the mass bias definitively as the number of BHs between these two models is fixed. Given the comparatively low number of BHs in the high mass range $>20\msun$, the increase in the rates by a factor of 2-3 suggests that a high percentage of these BHs will partake in a merger and represent a significant portion of total merging BHs.
\renewcommand{\arraystretch}{1.3}
\begin{table*}
\begin{tabular}{||@{}|c@{}|c@{}|c@{}|c@{}|c @{}|c @{}|c@{}| c@{}|c@{}||} 
 \hline
 disc & $\fform$ & BIMF & $\mathcal{R}_\rho$ & $\,\mathcal{R}_\rho$[5-20] & $\,\mathcal{R}_\rho$[20-50] & $\,\mathcal{R}_\rho$[50-100] & $\,\mathcal{R}_\rho$[20-50]/[5-20] & $\,\Gamma_\mathrm{LIGO}$ [yr$^{-1}$] \\%[0.5ex]
 \hline
SG$\alpha$ & Rowan & BIMF$_\text{Tagawa}^{\gamma=2.35}(\text{f})$ & 0.73  &  0.55    & 0.18 & -    & 0.33&  2.56\\
 %\hline
SG$\alpha$ & Rowan & BIMF$_\text{Tagawa}^{\gamma=1.7}$  & 3.92 & 2.26 & 1.65 &   -   & 0.73 & 16.9\\
SG$\alpha$ & Rowan & BIMF$_\text{Baxter}$              & 0.90 & 0.23 & 0.48 &  0.19 & 2.09 &  25.2\\
SG$\alpha$ & Rowan & BIMF$_\text{Bartos}^{\beta=2.5}$   & 0.80 & 0.34 & 0.40 &  0.06 & 1.18 &  12.5\\
SG$\alpha$ & Rowan & BIMF$_\text{Bartos}^{\beta=2.0}$   & 0.91 & 0.24 & 0.52 &  0.15 & 2.16 &  23.4\\

SG$\beta$ & Rowan & BIMF$_\text{Tagawa}^{\gamma=2.35}$ & 1.31  &  1.00    & 0.32 & -    & 0.32&  4.69\\
 %\hline
SG$\beta$ & Rowan & BIMF$_\text{Tagawa}^{\gamma=1.7}$  & 7.1 & 4.1 & 3.1 &   -   & 0.76 & 31.7\\
SG$\beta$ & Rowan & BIMF$_\text{Baxter}$              & 1.64 & 0.45 & 0.87 &  0.32 & 1.93 &  42.0\\
SG$\beta$ & Rowan & BIMF$_\text{Bartos}^{\beta=2.5}$   & 1.45 & 0.64 & 0.72 &  0.10 & 1.13 &  22.0\\
SG$\beta$ & Rowan & BIMF$_\text{Bartos}^{\beta=2.0}$   & 1.63 & 0.45 & 0.93 &  0.25 & 2.07 &  37.4\\

 && Range                            & 0.73-7.1 & 0.23-4.1 & 0.18-3.1 & 0.06-0.32 & 0.32-2.16 & 2.56-42.0 \\
 &&   Range [w.o BIMF$_\mathrm{Tagawa}^{\gamma=1.7}$]  & 0.73-1.64 & 0.23-0.64 & 0.18-0.93 & 0.06-0.32 & 0.32-2.16 & 2.56-42.0 \\
 &&  Observed                         &  17.9-44    & 13.3-39     &  2.5-6.8    &   0.1-0.4   &  0.09-0.29 &    \\ \hline
SG$\alpha$ & Tagawa & BIMF$_\text{Tagawa}^{\gamma=2.35}$ & 0.73  &  0.56    & 0.18 & -    & 0.32&  2.63\\
 %\hline
SG$\alpha$ & Tagawa & BIMF$_\text{Tagawa}^{\gamma=1.7}$  & 3.9 & 2.27 & 1.64 &   -   & 0.72 & 17.2\\
SG$\alpha$ & Tagawa & BIMF$_\text{Baxter}$              & 0.93 & 0.25 & 0.49 &  0.19 & 1.96 &  25.2\\
SG$\alpha$ & Tagawa & BIMF$_\text{Bartos}^{\beta=2.5}$   & 0.82 & 0.35 & 0.41 &  0.06 & 1.17 &  12.5\\
SG$\alpha$ & Tagawa & BIMF$_\text{Bartos}^{\beta=2.0}$   & 0.89 & 0.23 & 0.51 &  0.14 & 2.22 &  23.4\\
SG$\alpha$, $n_\mathrm{BH}=\text{const}$ & Rowan & BIMF$_\text{Tagawa}^{\gamma=2.35}$    & 0.79 & 0.56 & 0.21 & -    & 0.38 &  2.64\\
%\\[1ex]
 \hline
\end{tabular}
	\centering
	\caption{Table of results from Monte Carlo runs with different BIMFs and variations of initial conditions. \textit{From left to right:} The assumed disc viscosity model, the formation function, the BIMF function, the merger rate density $\mathcal{R}_\rho$ in Gpc$^{3}$yr$^{-1}$, the rate density in the binary mass range $5\msun\leq\mbin\leq20\msun$, density rate for $20\msun<\mbin\leq50\msun$, density rate for $50\msun<\mbin\leq100\msun$, ratio of density rates in second to first mass range, predicted detection rate of events for advanced LIGO, $\Gamma_\mathrm{LIGO}$. The (f) denotes the fiducial model which is used to test the merger rate change when the BH density is assumed to be constant (final row). The ranges of merger rates is shown in row 11/12 when BIMF$_\mathrm{Tagawa}^{\gamma=1.7}$ is/isn't included and compared to the currently available observed range from LIGO-VIRGO-KAGRA \protect\citep{Abbott2023} in row 13.}
	\label{tab:merger_rates}
\end{table*}
\subsection{Observable rates}
\label{sec:observable_rates}
The merger rate density $\mathcal{R}_\mathrm{\rho}$ in Gpc$^{-3}$yr$^{-1}$ is given by 
\begin{equation}
    \mathcal{R}_\mathrm{\rho}= \int \Gamma_\mathrm{AGN}(\mSMBH) \frac{dn_\mathrm{AGN}}{d\mSMBH}d\mSMBH\,.
    \centering
    \label{eq:AGN_mass_funct2}
\end{equation}
For our fiducial model BIMF$_\mathrm{Tagawa}^{\gamma=2.35}$ (SG$\alpha$), this results in a rate of $0.73$Gpc$^{-3}$yr$^{-1}$, more than an order of magnitude lower than the value from current LVK data. 

To calculate the rate of BH mergers per year from Earth, we adopt the observational horizon distance $D_\mathrm{h}$ for Advanced LIGO for which a binary with mass $\mbin$ is detectable at a signal to noise ratio of 8 \citep[see][]{Dominik2015}
\begin{equation}
    \centering
    D_\mathrm{h}(\mbin) = 0.45\bigg(\frac{\mbin}{2.8\msun}\bigg)^{5/6}\text{Gpc}\,.
    \label{eq:observable_distance}
\end{equation}
The comoving volume $V_\mathrm{c}$ in which we can detect a merger with binary mass $\mbin$ is given by
\begin{equation}
    V_\mathrm{c}(\mbin)= \frac{4}{3}\pi\bigg(\frac{D_\mathrm{h}(\mbin)}{2.26}\bigg)^{3}(1+z)^{-3}\,,
    \centering
    \label{eq:AGN_mass_funct3}
\end{equation}
where $z$ is redshift, that we neglect for the purpose of this study and set to zero. The observed rate $\Gamma_\mathrm{LIGO}$ per year is then
\begin{equation}
    \centering
    \Gamma_\mathrm{LIGO} = \iint V_\mathrm{c}(\mbin)\frac{dR_\mathrm{AGN}(\mbin,\mSMBH)}{d\mbin}\frac{dn_\mathrm{AGN}}{d\mSMBH}d\mbin\,d\mSMBH\,.
    \label{eq:LIGO_rate}
\end{equation}
The merger rate per binary mass term, $dR_\mathrm{AGN}/d\mbin$ is evaluated using the merging binary mass distribution from each $\mSMBH$ value put through our model. For lower $\mSMBH$ masses that have a low merger number, we repeat the analysis until we have a merger sample of at least 1000 BHs. We display the merger rate densities, LIGO merger rate and the merger rates for binary masses in the ranges $5\leq\mbin\leq20$, $20<\mbin\leq50$ and $50<\mbin\leq100$
in Table \ref{tab:merger_rates} for all three BIMFs and varying initial conditions. The varied initial conditions include: maintaining a fixed BH density over $t_{\rm AGN}$ and a rerun of the entire SG$\alpha$ suite assuming the binary formation criterion of \citet{Tagawa2020} (Eq. \ref{eq:fform_tag}). 

Our models give us a merger rate of $\Gamma_\mathrm{LIGO}$=2.56-42.0yr$^{-1}$ with a local density rate of $0.73\,-\,7.1$Gpc$^{-3}$yr$^{-1}$. While not within the range of 17.9--44Gpc$^{-3}$yr$^{-1}$ from LIGO-VIRGO-KAGRA \citep{Abbott2023}, it suggests the contribution from the AGN channel is non-negligible. The merger rate increases by approximately a factor 2 going from the SG$\alpha$ to SG$\beta$ discs. This is a direct result of an increased gas density in the inner radial regions of $\mSMBH=10^{7}\msun-10^{9}\msun$ (see Figure \ref{fig:C_prof}), which reduces the alignment time (Eq. \ref{eq:t_align_full}) and the effective formation timescale. We find the merger rate is \textit{highly} dependent on the number of BHs and the most sensitive parameter to the rates across the simulations in this work. Hence, we encourage further constraint of the expected number of BHs within $R_\mathrm{inf}$ in the Universe. When larger initial BH masses are permitted in the initial distribution, as in BIMF$_\mathrm{Baxter}$, the rates become dominated by larger binary masses ($\mbin>20\msun$), despite BHs of masses $>10\msun$ being fewer in number. This result remains consistent across the SG$\alpha$, SG$\beta$ models and the two different formation functions. This hardening of the merging mass function in AGN was found in \citet{Yang2019}, in the context of pre-existing binary mergers, driven by migration traps and mass biased alignment time. Here, we again have the biased alignment time ($t_\mathrm{align}\sim \mBH^{-1}$, Eq. \ref{eq:t_align_full}). In addition to this, larger BHs have a considerably shorter BBH formation timescale since their Hill radius is larger and $\langle \fform \rangle_\mathrm{M}$ is far larger. Hence high mass mergers could make up a significant fraction of the anticipated rates in the AGN channel, as shown by Table~\ref{tab:merger_rates}.

The values of $\Gamma$ and $\mathcal{R}_\rho$ as a function of $\mSMBH$ for our SG$\alpha$, SG$\beta$ and test suite with the $f_\mathrm{form,Tag}$ and SG$\alpha$ assumption are compared in Figure \ref{fig:rates_over_M_SMBH}. Between the two disc models, the SG$\beta$ discs show an increased merger rate across $\mSMBH$, as anticipated from the lower viscosity and higher density. The merger rates diverge further from the SG$\alpha$ models as $\mSMBH$ increases, stemming from the boundary of the radiation zone ($\beta\ll 1$) moving further out in $R$, covering a larger portion of the domain. The rates calculated assuming $f_\mathrm{form, Tag}$ remain nearly identical to the original SG$\alpha$ models across the range in $\mSMBH$.
\subsection{Merger properties}
\label{sec:merger_props}
\begin{figure*}
    \centering
    \includegraphics[width=17cm]{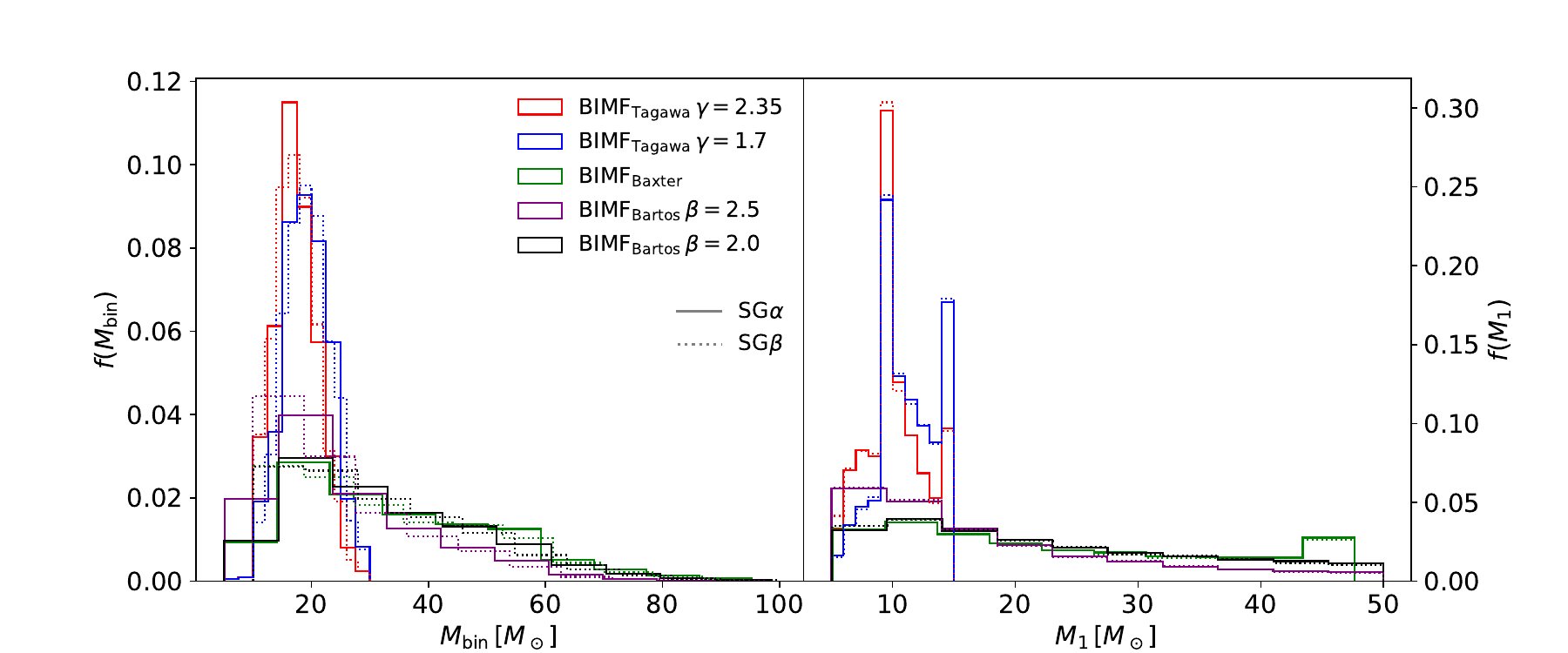}
    \caption{\textit{Left:} Mass distribution of the merging binary mass $\mbin$ for each BIMF outlined in Sec.~\ref{sec:pop_sample}, represented by different colours. \textit{Right:} Mass distribution of the primary black hole mass $M_1$ of mergers. The results show a significant hardening of the merging BH mass function compared to the BIMF, indicating larger black holes have a much greater chance to form binaries and merge.}
    \label{fig:mass_distribution}
\end{figure*}

\begin{figure}
    \centering
    \includegraphics[width=8cm]{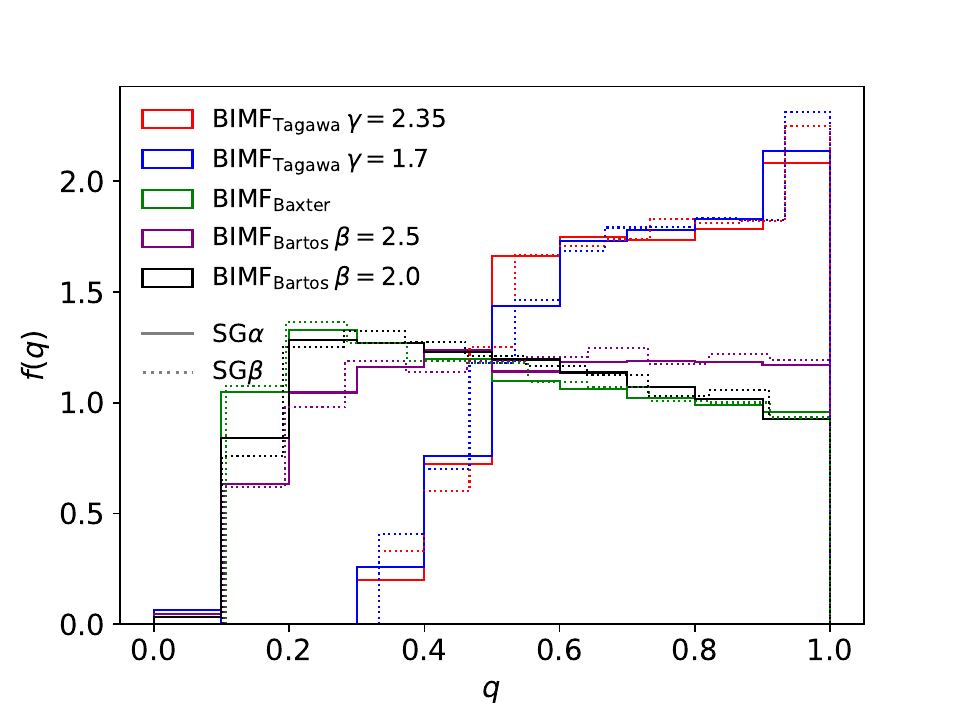}
    \caption{Mass ratio distribution $q=M_2/M_1$ for our merging binaries for each BIMF (colour coded). Demonstrating the AGN channel can easily produce high mass ratio mergers.}
    \label{fig:q_distribution}
\end{figure}

\begin{figure}
    \centering
    \includegraphics[width=8cm]{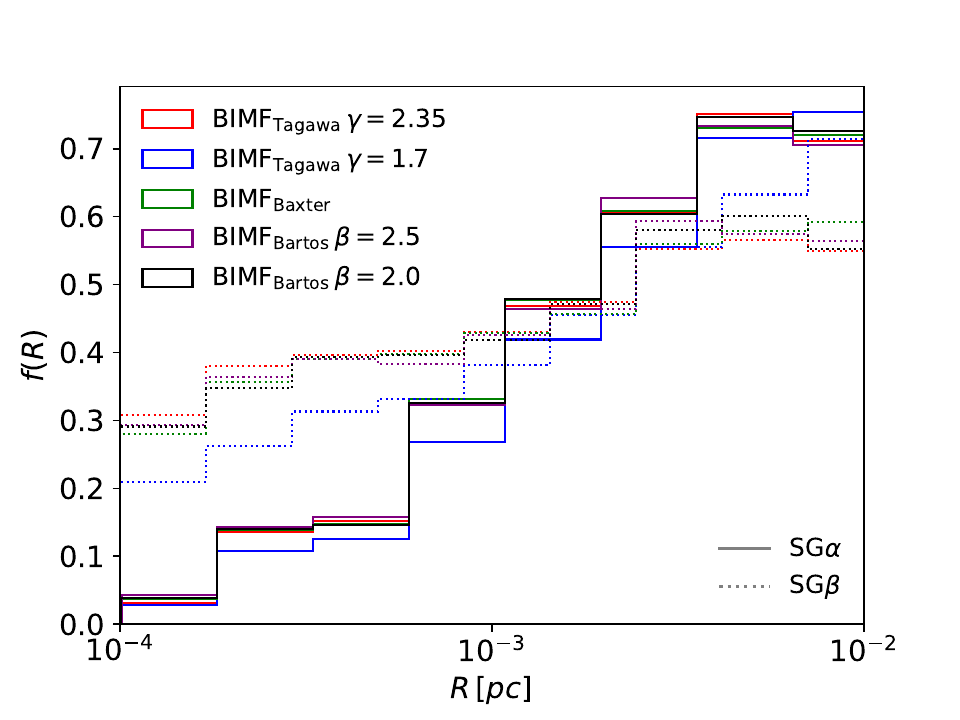}
    \caption{Distribution of radial positions in the disc $R$ for our merging binaries for each BIMF (colour coded). Merger rate peaks is greatest around the outer radial bound of the simulationsm, outside of which the AGN disc becomes gravitationally unstable under the Sirko-Goodman model.}
    \label{fig:R_distribution}
\end{figure}

We show the distributions of mass, mass ratio and radial position of merging binaries generated by our model averaged over all $\mSMBH$, weighted by the AGN mass function in Figures \ref{fig:mass_distribution}, \ref{fig:q_distribution} and \ref{fig:R_distribution} respectively.
The mass distribution reaffirms the aforementioned mass hardening effect of the AGN channel. For example, consider BIMF$_\mathrm{Baxter}$, which demonstrates a near flat profile in $M_1$ compared to the original profile $\sim\mBH^{-2.2}$. Considering the two simple power laws of BIMF$_{\rm Bartos}$, we find the exponent for the power law distribution of $M_{1}$ is flattened by a factor $\Delta \xi\simeq1.1-1.2$ compared to the  BIMF, i.e $M_1\propto \mBH^{\xi_{0} + \Delta \xi}$, where $\xi_0$ is the exponent of the BIMF. This relatively good agreement with the value $\Delta \xi\simeq1.3$ predicted by \citet{Yang2019}. This is also reflected in the $q$ distribution. For flatter BIMFs with higher limits on $\mBH$ (i.e BIMF$_\mathrm{Baxter}$,BIMF$_\mathrm{Bartos}$), the $q$ distribution becomes less steep and can even flip to favour higher mass ratio mergers, consistent with the recent findings of \citet{Delfavero2024} who account for additional physics (migration, repeated mergers). Qualitatively, this means less numerous high mass BHs can very easily encounter an merge with abundant low mass BHs. This has optimistic prospects for second (or more) generation mergers as the merged BH can more easily repeat the formation and merger process thanks to its larger mass, potentially further hardening the merging mass distribution. Such a scenario could easily explain massive and or comparable component-mass binaries such as GW190814 and GW190521 \citep[e.g][]{Abbott2020_GW190521,Abbott2020}. Comparing the rates in the three mass bins of Table \ref{tab:merger_rates}, we find mergers in AGN may account for $1.7-40\%$ mergers in the range $5\msun<\mbin\leq20\msun$, $\sim3-100\%$ mergers in the range $20\msun<\mbin\leq50\msun$ and $\sim15-100\%$ of mergers in $20\msun<\mbin\leq50\msun$
Therefore, AGN may not be the primary source of observed low mass BH mergers, but could account for a large fraction of high mass mergers. Hence, the relative contribution of the AGN channel could potentially be constrained using the ratio of merger rates in different mass bins. Using the ratio between the first and second mass bins (Table \ref{tab:merger_rates}), we find a preference for a higher ratio in the AGN channel than the observed rate. This would again indicate a large contribution to the rates from other channels for low mass mergers with a steeper mass distribution. As more GW detections are made and the BIMF in AGN better constrained, it could then be possible to constrain more rigorously the relative contribution from the AGN channel from the merging mass distribution.

The radial distribution of mergers (Figure \ref{fig:R_distribution}) varies less than an order of magnitude across $R$, peaking at the outer radial bound of the simulations. At lower radii, we are limited by a low number of BHs in each cylindrical volume and increased Keplerian shear $\sigma_\mathrm{Kep}$. The positive slope of the curve is consistent with \citet{Tagawa2020}, who find the majority of gas-capture encounters occur in the outer regions of the AGN disc. Note we have neglected the possibility of repeated mergers and radial migration, which we expect will change this radial profile in reality.
\section{Summary and Conclusions}
\label{sec:conclusions}
In this work, we predicted the merger rate of BBHs in AGN formed via BH-BH scatterings in an AGN disc using the physically motivated formation prescription based on \citet{Rowan2023} \citep[see also][]{Whitehead2023}. The primary goal was to test whether implementing the prescription, derived from high resolution fully hydrodynamical simulations, alters the rates significantly compared to simplified dynamical friction models. Using a range of initial black hole mass functions, we constrain the merger rate density to $0.73-7.1$Gpc$^{-3}$pc$^{-1}$. These rates corroborate the range predicted from many analytical studies \citep[e.g][]{Bartos2017,Yang2019,Tagawa2020,McKernan2020,Delfavero2024}. We note that \cite{Delfavero2024} find the merger rate peaks at higher $M_\bullet$ and posit this could stem from their assumption that BHs that pass within $r_\mathrm{H}$ will reliably become bound independently of $\mSMBH$, where we find the formation function drops by approximately an order of magnitude in the range $10^{7}-10^{9}\msun$. Therefore we encourage more detailed studies in the future to include a formation function when simulating BH scatterings in AGN discs.

We find that the rates are very similar (within $10\%$) when the gas-capture process is modelled using an analytic dynamical friction treatment of the gas-assisted binary formation process, affirming the results of earlier studies that use this simplified model. 

The mass distribution (both the binary and primary mass) of merging BHs is significantly more top heavy compared to the initial BH mass distribution, due to a favourability for high mass BHs to align with the disc and form binaries via gas-dissipation. Therefore the merger rates and masses are sensitive to the assumed BIMF, where more top heavy BIMFs lead to increases in the rates. This bias also leads to a more flat mass ratio distribution, thus the AGN channel can easily explain the high mass and high mass ratio detections from GW observatories. Mergers are more numerous closer to the gravitationally unstable region of the AGN disc (higher $R$), although the distribution becomes flatter when a pressure dependent disc viscosity is assumed. We find the overall merger rate to lie within $\lesssim5-40\%$ that of LIGO-VIRGO-KAGRA, suggesting a non-negligible contribution from the AGN channel. However, due to the top heavy merging mass function, the AGN channel can potentially be the source of $\sim3-100\%$ mergers in the range $20\msun<\mbin\leq50\msun$ and $\sim15-100\%$ of mergers in $50\msun<\mbin\leq100\msun$. Due to the mass bias of the gas-assisted binary formation mechanism, we encourage future studies to account for repeated mergers, which may further increase the merger rate for $\mbin>30\msun$. We posit that reductions in the observation merger rate uncertainty and better constraints of the BIMF in AGN could allow us to constrain the relative contribution from AGN using the relative merger rates from low and high mass binaries.

Though we have used a well motivated formation criterion for our binaries, our model is still subject to several large assumptions. The formation function assumes that the hydrodynamics of the encounter are isothermal. Our recent 2D non-isothermal hydrodynamical work \citet{Whitehead2023_novae} has shown that this is an oversimplification, where gas heating can be significant during the encounter, reducing the gas mass in the Hill sphere and potentially reducing the typical value of $\fform$. We have pessimistically assumed all BHs are formed in the spherical stellar cluster and the merging systems must first align with the disc. In reality the parent stars could align first (and faster) before the BH is born \citep{Panamarev2018}. We assume there is no migration in the disc, which could speed up the rate of encounters. Binary-single scatterings with a tertiary BH or star that could potentially ionise the BBH or induce a merger are also ignored, alongside BBH formations from three-body scatterings.

The results indicate that the number of BHs produced/present around the AGN disc is the most influential parameter on the BH rates and therefore encourage future projects to constrain this value and its possible dependence on $\mSMBH$ itself. We conclude that BHs merging in AGN through the gas capture mechanism are a non-negligible contributor to the observed rates and a potentially dominant channel for high mass and comparable binary component mass mergers. 

\section*{Acknowledgements}
\begin{itemize}
    \item This simulations which were used to construct the formation function were performed using the Cambridge Service for Data Driven Discovery (CSD3), part of which is operated by the University of Cambridge Research Computing on behalf of the STFC DiRAC HPC Facility (www.dirac.ac.uk). The DiRAC component of CSD3 was funded by BEIS capital funding via STFC capital grants ST/P002307/1 and ST/R002452/1 and STFC operations grant ST/R00689X/1. DiRAC is part of the National e-Infrastructure.
    \item  This work was supported by the Science and Technology Facilities Council Grant Number ST/W000903/1. 
\end{itemize}
\bibliographystyle{mnras}
\bibliography{Paper} 
% do not change these lines
%\bsp	% typesetting comment
\label{lastpage}
\end{document}